%% file: quant_exptdes07.tex
\documentclass[12pt]{article}
\input{articlelayout}
\input{rlktexdefs}

\usepackage{epsfig}
\usepackage{times}

\usepackage[dvips]{color}
\newcommand{\red}[1]{\textcolor{red}{#1}}
\newcommand{\blue}[1]{\textcolor{blue}{#1}}
\newcommand{\green}[1]{\textcolor{green}{#1}}

\newcommand{\black}[1]{\textcolor{black}{#1}}

\usepackage{psfrag}

\newcommand{\epsfmaxlike}[1]
{\epsfbox{#1}}

\newcommand{\epsfboxit}[1]
{\epsfbox{#1}}

\newcommand{\surr}{{\rm surr}}
\newcommand{\rhosurr}{\rho^\surr}
\newcommand{\fsurr}{f^\surr}
\newcommand{\Xsurr}{X^\surr}
\newcommand{\qsurr}{q^\surr}

\renewcommand{\refsec}[1]{\S \ref{sec:#1}}

\newcommand{\wc}{{\rm wc}}
\newcommand{\Ceq}{C_{\rm eq}}

\newcommand{\mcol}[3]{\multicolumn{#1}{#2}{#3}}

\newcommand{\init}{{\rm init}}

\newcommand{\nexsub}{\ell^{\rm sub}_{\rm expt}}
\newcommand{\round}[1]{ {\bf round}\left\{#1\right\} }

\newcommand{\btab}{\begin{tabular}}
\newcommand{\etab}{\end{tabular}}

\newcommand{\Kb}{\bar{K}}

\newcommand{\pih}{\hat{\pi}}
\newcommand{\pit}{\pi^{\rm true}}

\newcommand{\blues}[1]{\blue{#1^\surr}}

\newcommand{\thtrue}{\th^\true}
\newcommand{\nth}{{n_\th}}
\newcommand{\sumalf}{\sum_{\alf}}

\newcommand{\sumgam}{\sum_{\gam}}

\newcommand{\true}{{\rm true}}
\newcommand{\ptrue}{p^{\rm true}}
\newcommand{\sigtrue}{\sig^{\rm true}}
\newcommand{\pemp}{p^{\rm emp}}

\newcommand{\zh}{\hat{z}}

\newcommand{\crao}{Cram\'{e}r-Rao\ }
\newcommand{\kapgam}{{\kappa_\gam}}

\newcommand{\rank}{{\bf rank}}
\newcommand{\qsys}{Q}

\newcommand{\lamopt}{\lam^{\rm opt}}
\newcommand{\lamsub}{\lam^{\rm sub}}
\newcommand{\lamopto}{\lamopt_\caseo}
\newcommand{\lamoptt}{\lamopt_\caset}

\newcommand{\thml}{\th^{\rm ML}}
\newcommand{\rhoml}{\rho^{\rm ML}}
\newcommand{\ftrue}{f^{\rm true}}
\newcommand{\rhotrue}{\rho^{\rm true}}
\newcommand{\psitrue}{\psi_{\rm true}}
\newcommand{\rhopur}{\rho_{\rm pure}}
\newcommand{\rhomix}{\rho_{\rm mixd}}

\newcommand{\caseo}{\pur}
\newcommand{\caset}{\mix}
\newcommand{\pur}{{\rm pure}}
\newcommand{\mix}{{\rm mixd}}

\newcommand{\nin}{{n_{\rm in}}}
\newcommand{\nout}{{n_{\rm out}}}
\newcommand{\noutb}{{\bar{n}_{\rm out}}}
\newcommand{\nconf}{{n_{\rm cfg}}}

\newcommand{\nr}{n_{\rm sub}}

\renewcommand{\lamh}{\lam^{\rm opt}}

\newcommand{\lamr}{\lam^{\rm sub}}

\newcommand{\pcyn}{\nu_{1|0}}
\newcommand{\pcny}{\nu_{0|1}}
\newcommand{\pcnn}{\nu_{0|0}}
\newcommand{\pcyy}{\nu_{1|1}}

\newcommand{\pny}{(1-\eta)(1-\del)}

\newcommand{\Ubell}{U_{\rm bell}}

\newcommand{\Uswap}{U_{\rm swap}}
\newcommand{\Usqsw}{\sqrt{\Uswap}}

\newcommand{\had}{ {\rm had} }

\newcommand{\ellh}{\hat{\ell}}
\newcommand{\ellrnd}{\ell^{\rm round}_{\rm expt}}
\newcommand{\omh}{\hat{\om}}
\newcommand{\avlike}{\avg\ L}
\newcommand{\Udes}{U_{\rm des}}
\newcommand{\Uact}{U_{\rm act}}

\newcommand{\iti}{ {(i)} }
\newcommand{\itip}{ {(i+1)} }

\newcommand{\tf}{{t_{\rm f}}}

\newcommand{\Mb}{\bar{M}}

\renewcommand{\vec}{{\bf vec}}

\newcommand{\cov}{{\bf cov}}
\newcommand{\var}{{\bf var}}

\newcommand{\Th}{\Theta}

\newcommand{\nex}{{\ell_{\expt}}}
\newcommand{\nsa}{{n_{\rm sa}}}

\newcommand{\Cbfnn}{\Cbf^{n\times n}}
\newcommand{\Cbfntnt}{\Cbf^{n^2\times n^2}}

\renewcommand{\sett}[2]{ \left\{ \,#1\, \Big| \,#2\, \right\} }

\newcommand{\Uhad}{U_{\rm had}}

\renewcommand{\norm}[1]{ \| #1 \| }
\renewcommand{\bar}[1]{\overline{#1}}

\renewcommand{\hat}[1]{\widehat{#1}}

\renewcommand{\rhoh}{\hat{\rho}}

\renewcommand{\mbf}[1]{\mbox{\boldmath $#1$}}

\newcommand{\ket}[1]{\mbf{|}#1\mbf{\rangle}}
\newcommand{\bra}[1]{\mbf{\langle}#1\mbf{|}}

\newcommand{\avg}{ {\bf E} }
\newcommand{\trace}{{\rm Tr}}

\renewcommand{\hbar}{ {h {\!\!\!^{\scriptscriptstyle -} } } }

\newcommand{\ketlo}{\ket{0}}
\newcommand{\kethi}{\ket{1}}
\newcommand{\ketp}{\ket{{\scriptstyle +}}}
\newcommand{\ketm}{\ket{{\scriptstyle -}}}

\newcommand{\prob}[1]{ {\bf Prob}\left\{ #1 \right\} }

\newcommand{\mattwo}[4]{
\left[
\bea{cc}
\ds #1 & \ds #2 \\ \ds #3 & \ds #4
\eea
\right]
}

\begin{document}


\title{
Optimal Experiment Design\\ for\\ 
Quantum State and Process Tomography
\\ and\\
Hamiltonian Parameter Estimation
\footnote{Research supported by the DARPA QUIST Program.}
}

\author{
Robert L. Kosut\thanks{
SC Solutions, Sunnyvale, CA, USA,
{\tt kosut@scsolutions.com}
}
\and
Ian Walmsley\thanks{
Oxford University,
Oxford, UK,
{\tt walmsley@physics.ox.ac.uk}
}
\and
Herschel Rabitz\thanks{
Princeton University,
Princeton, NJ,
{\tt hrabitz@princeton.edu}
}
}

\date{}

\maketitle
\thispagestyle{empty}
\begin{abstract}

A number of problems in quantum state and system identification are
addressed. Specifically, it is shown that the maximum likelihood
estimation (MLE) approach, already known to apply to quantum state
tomography, is also applicable to quantum process tomography
(estimating the Kraus operator sum representation (OSR)), Hamiltonian
parameter estimation, and the related problems of state and process
(OSR) distribution estimation. Except for Hamiltonian parameter
estimation, the other MLE problems are formally of the same type
of convex optimization problem and therefore can be solved very
efficiently to within any desired accuracy.

Associated with each of these estimation problems, and the focus of
the paper, is an optimal experiment design (OED) problem invoked by
the \crao Inequality: find the number of experiments to be performed
in a particular system configuration to maximize estimation accuracy;
a configuration being any number of combinations of sample times,
hardware settings, prepared initial states, \etc. We show that in all
of the estimation problems, including Hamiltonian parameter
estimation, the optimal experiment design can be obtained by solving a
convex optimization problem.\footnote{
Software to solve the MLE and OED convex optimization problems is
available upon request from the first author.
}

\end{abstract}

\newpage

\tableofcontents

\newpage
\pagestyle{plain}
\pagenumbering{arabic}

\section{Introduction}
\label{sec:intro}

\begin{quote}
\begin{small}
{\em ``In a machine such as this {\em [a quantum computer]} there are
very many other problems due to imperfections. For example, in the
registers for holding the data, there will be problems of cross-talk,
interactions between one atom and another in that register, or
interaction of the atoms in that register directly with things that
are happening along the program line that we didn't exactly bargain
for. In other words, there may be small terms in the Hamiltonian
besides the ones we've written. Until we propose a complete
implementation of this, it is very difficult to analyze. At least some
of these problems can be remedied in the usual way by techniques such
as error correcting codes and so forth, that have been studied in
normal computers. But until we find a specific implementation for this
computer, I do not know how to proceed to analyze these effects.
However, it appears that they would be very important in
practice. This computer seems to be very delicate and these
imperfections may produce considerable havoc.''}

-- Richard P. Feynman,
``Quantum Mechanical Computers,''
{\em Optics News}, February 1985.

\end{small}
\end{quote}

\subsection{Alleviating the ``havoc''}
\label{sec:intro havoc}

The concerns heralded by Feynman remain of concern today in all the
implementations envisioned for quantum information systems. In a
quantum computer it is highly likely that in order to achieve the
desired system objectives, these systems will have to be tuned, or
even entirely determined, using estimated quantities obtained from
data from the actual system rather than solely relying on an initial
design from a theoretical model. The problem addressed here is to
design the experiment in order to yield the optimum information for
the intended purpose. This goal is not just limited to quantum
information systems. It is an essential step in the engineering
practice of system identification \cite[Ch.14]{Ljung:87}.  That is,
the design of the experiment which gives the best performance against
a given set of criteria, subject to constraints reflecting the
underlying properties of the system dynamics and/or costs associated
with the implementation of certain operations or controls.

Clearly each application has a specific threshold of performance.
For example, the requirements in quantum chemistry are generally not
as severe as in quantum information systems.  The objective of a
measurement, therefore, depends on the way in which information is
encoded into the system to begin with, and this is in turn, depends on
the application. In this paper we are concerned with estimating
quantum system properties: the state, the process which transforms the
state, and parameters in a Hamiltonian model.

The estimation of the state of a quantum system from available
measurements is generally referred to as {\em quantum state
tomography} about which there is extensive literature on both
theoretical and experimental aspects, \eg, see
\cite[Ch.8]{NielsenC:00}, \cite{GriceW:96} and the references therein.
The more encompassing procedure of {\em quantum system identification}
is not so easily categorized as the nomenclature (and methodology)
seems to depend on the type and/or intended use of the identified
model. For example, {\em quantum process tomography} (QPT) refers to
determining the Kraus {\em operator-sum-representation} (OSR) of the
input state to output state (completely positive) map, \eg,
\cite[\S 8.4.2]{NielsenC:00}, \cite{ChuangN:97}.  {\em Hamiltonian
parameter estimation} refers to determining parameters in a model of
the system Hamiltonian, \eg, \cite{Mabuchi:96}, \cite{ChildsPR:00},
\cite{GeremiaR:02}, \cite{VDM:01}. Somewhere in between quantum
process tomography and Hamiltonian parameter estimation is {\em
mechanism identification} which seeks an estimate of population
transfer between states as the system evolves, \eg, \cite{MitraR:03}.

{\em Maximum likelihood estimation} (MLE), a well established method
of parameter estimation which is used extensively in current
engineering applications, \eg, \cite{Ljung:87}, was proposed in
\cite{Banaszek:98,ParisETAL:01} and \cite{Sacchi:01} for quantum state
tomography of a quantum system with non-continuing measurements, \ie,
data is taken from repeated identical experiments.  Also, as observed
in \cite{ParisETAL:01,Sacchi:01}, the MLE of the density matrix is a
convex optimization problem.

In this paper we address the related problem of optimal experiment
design (OED) so as to secure an estimate of the best quality. The
approach presented relies on minimizing the Cramer-Rao lower bound
\cite{Cramer:46} where the design parameters are the number of
experiments to be performed while the system is in a specified
configuration. En route we also show that many related problems in
state and process tomography can also be solved using MLE, and
moreover, they are all {\em formally} the same type of convex
optimization problem, namely, a determinant maximization problem,
referred to as a {\em maxdet problem}
\cite{BoydV:04,VBW:98}. Similarly, the OED problem posed here is also
of a single general type of convex optimization problem, namely, a
{\em semidefinite program} (SDP).

Convexity arises in many ways in quantum mechanics and this is briefly
discussed in \refsec{cvx}.  The great advantage of convex optimization
is a globally optimal solution can be found efficiently and reliably,
and perhaps most importantly, can be computed to within any desired
accuracy. Achieving these advantages, however, requires the use of
specialized numerical solvers. As described in \refsec{software}, the
appropriate convex solvers have been embedded in some software tools
we have composed which can solve the MLE and OED problems presented
here. 

In the remainder of the paper we present both MLE and the
corresponding OED as applied to: quantum state tomography (MLE in
\refsec{mle} and OED in \refsec{expt des}), estimating the
distribution of known input states (MLE in \refsec{dist est} and OED
in \refsec{expdes qs dist}), quantum process tomography using the
Kraus operator sum representation (MLE in \refsec{qpt osr} and
OED in \refsec{expdes osr}), estimating the distribution of a known
OSR set (MLE in \refsec{osr dist} and OED in \refsec{expdes osr
dist}), and to Hamiltonian parameter estimation (MLE in \refsec{mle
hampar} and OED in \refsec{ham oed}). A summary in table form is
presented in \refsec{summary} followed by a discussion in \refsec{iter
adapt} of the relation of MLE and OED to iterative adaptive control of
quantum systems.

\subsection{Convexity and quantum mechanics}
\label{sec:cvx}

Many quantum operations form convex sets or functions. Consider, for
example, the following convex sets which arise from some of the basic
aspects of quantum mechanics:
\bc
\begin{tabular}{lll}
\begin{tabular}{l}
probability outcomes
\end{tabular}
&
$\seq{p_\alf\in\Rbf}$
&
$
\sum_\alf p_\alf=1,
\;\;\;
p_\alf\geq 0
$
\\&&\\
\begin{tabular}{l}
density matrix
\end{tabular}
&
$\seq{\rho\in\Cbfnn}$
&
$
\trace\ \rho=1,
\;\;\;
\rho \geq 0
$
\\&&\\
\begin{tabular}{l}
positive operator
\\ 
valued measure (POVM)
\end{tabular}
&
$\seq{O_\alf\in\Cbfnn}$
&
$
\sum_\alf\ O_\alf = I_n,
\;\;\;
O_\alf \geq 0
$
\\&&\\
\begin{tabular}{l}
operator sum 
\\
representation (OSR)
\\
in fixed basis
\\ 
$\set{B_i\in\Cbfnn}{i=1,\ldots,n^2}$
\end{tabular}
&
$\seq{X\in\Cbf^{n^2\times n^2}}$
&
$
\sum_{ij}\ X_{ij}\ B_i^*B_j = I_n,
\;\;\;
X \geq 0
$
\end{tabular}
\ec
An example of a convex function relevant to quantum information is
{\em worst-case gate fidelity}, a measure of the ``distance'' between
two unitary operations on the same input. As pointed out in
\cite{GilchristETAL:04}, there are many ways to define this
measure. Consider, for example,
\beq[eq:wc fidelity]
f^\wc(\Udes,\ \Uact)
=
\min_{\norm{\psi}=1}
\left|
\left(\Udes \psi\right)^*
\left( U_{\rm act}\psi \right)
\right|^2
\eeq
where $\Udes\in\Cbfnn$ is the desired unitary and $U_{\rm
act}\in\Cbfnn$ is the actual unitary. In this case the worst-case
fidelity can be interpreted as the minimum probability of obtaining
the desired output state $\Udes\psi$ over all possible pure input
states $\psi$ which produce the actual output state $\Uact\psi$.  If
$\Udes$ and $\Uact$ differ by a scalar phase then the worst-case
fidelity is clearly unity; which is consistent with the fact that a
scalar phase cannot be measured.  This is not the case for the error
norm $\norm{\Udes-\Uact}$.  

As shown in Appendix \refsec{fidelity}, obtaining the worst-case
fidelity requires solving the following (convex) {\em quadratic
programming} (QP) problem:
\beq[eq:fidelity qp]
\bea{ll}
\mbox{minimize}
&
z^T(aa^T+bb^T)z
\\
\mbox{subject to}
&
\sum_{k=1}^n\ z_k = 1,
\;\;\;
z_k \geq 0
\eea
\eeq
with the vectors $a,b$ in $\Rbf^n$ the real and imaginary parts,
respectively, of the eigenvalues of the unitary matrix $\Udes^*\Uact$,
that is,
$a=\real\ {\bf eig}(\Udes^*\Uact),\ 
b=\imag\ {\bf eig}(\Udes^*\Uact)$.
In some cases it is possible to compute the worst-case fidelity
directly, \eg, in the example in Section \refsec{ex hampar} and in
some examples in \cite[\S 9.3]{NielsenC:00}.  Although the optimal
objective value $f^\wc(\Udes,\ \Uact)$ is global, the optimal
worst-case state which achieves this value is not unique.

In addition to these examples, convex optimization has been exploited
in \cite{AudD:03} and \cite{Sacchi:01} in an attempt to realize
quantum devices with certain properties. In \cite{EldarMV:03} and
\cite{KosutWER:04}, convex optimization is used to design optimal
state detectors which have the maximum efficiency.

In general, convex optimization problems enjoy many useful properties.
From the introduction in \cite{BoydV:04}, and as already stated, the
solution to a convex optimization problem can be obtained to within
any desired accuracy. In addition, computation time does not explode
with problem size, stopping criteria always produce a lower bound on
the solution, and if no solution can be found a proof of infeasibility
is provided. There is also a complete duality theory which can yield
more efficient computation as well as optimality conditions. This is
explored briefly in Section \refsec{dual}.

\subsection{Software for tomography \& experiment design}
\label{sec:software}

We have composed some MATLAB m-files which can be used to solve a
subset of the QPT and OED convex optimization problems presented
here. The examples shown here were generated using this software.  The
software, available upon request from the first author, requires the
convex solvers YALMIP \cite{Yalmip} and SDPT3 \cite{Sdpt3} which can
be downloaded from the internet. These solvers make use of {\em
interior-point methods} for solving convex optimization problems, \eg,
\cite[Ch.11]{BoydV:04}, \cite{NN94}.


\section{Quantum State Tomography}
\label{state est}

Consider a quantum system which has $\nout$ distinct {\em outcomes},
labeled by the index $\alf,\ \alf=1,\ldots,\nout$, and which can be
externally manipulated into $\nconf$ distinct {\em configurations},
labeled by the index $\gam,\ \gam=1,\ldots,\nconf$.  Configurations
can include wave-plate angles for photon counting, sample times at
which measurements are made, and settings of any experimental
``knobs'' such as external control variables, \eg, laser wave shape
parameters, magnetic field strengths, and so on. For quantum process
tomography (\refsec{qpt osr}) and Hamiltonian parameter estimation
(\refsec{mle hampar}), configurations can also include distinctly
prepared initial states.

The problem addressed in this section is to determine the minimum
number of experiments per configuration in order to obtain a state
estimate of a specified quality, \ie, what is the tradeoff between
number of experiments per configuration and estimation quality. The
method used to solve this problem is based on minimizing the size of
the Cram\'{e}r-Rao lower bound on the estimation error
\cite{Cramer:46}.

\subsection{Data collection}
\label{sec:data collect}

\begin{figure}[h]
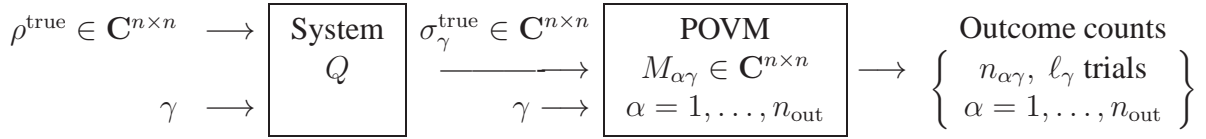

\[
\bea{rr}
\rhotrue\in\Cbfnn & \longrightarrow
\\
\mbox{} & \mbox{}
\\
\gam & \longrightarrow
\eea
\mathbox{
\bea{c}
\mbox{System}
\\
\qsys
\\
\mbox{}
\eea
}
\bea{r}
\sigtrue_\gam\in\Cbfnn 
\\ 
\mbox{----------}\!\!\!\longrightarrow 
\\ 
\gam \longrightarrow
\eea
\mathbox{
\bea{c}
\mbox{POVM}
\\
M_{\alf\gam} \in\Cbfnn
\\
\alf=1,\ldots,\nout
\eea
}
\longrightarrow
\bea{c}
\mbox{Outcome counts}
\\
\seq{
\bea{c} n_{\alf\gam},\ \ell_\gam~\mbox{trials} 
\\ \alf=1,\ldots,\nout \eea
}
\eea
\]
\caption{System/POVM.}
\label{fig:rho}
\end{figure}

The data is collected using a procedure referred to here as {\em
non-continuing measurements}.  Measurements are recorded from
identical experiments in each configuration $\gam$ repeated
$\ell_\gam$ times.  The set-up for data collection is shown
schematically in Figure \ref{fig:rho} for configuration $\gam$.

\noindent
Here $\rhotrue\in\Cbfnn$ is the true, unknown state to be estimated,
$\sigtrue_\gam\in\Cbfnn$ is the {\em reduced density matrix} which
captures all the statistical behavior of the $\qsys$-system under the
action of the measurement apparatus, and $n_{\alf\gam}$ is the number
of times outcome $\alf$ is obtained from the $\ell_\gam$ experiments.
Thus,
\beq
\sumalf\ n_{\alf\gam} = \ell_\gam,
\;\;
\nex=\sumgam\ \ell_\gam
\eeq
where $\nex$ is the total number of experiments.  The {\em data set}
consists of all the outcome counts,
\beq[eq:data]
D
=
\set{n_{\alf\gam}}
{\alf=1,\ldots,\nout,\gam=1,\ldots,\nconf}
\eeq
The design variables used to optimize the experiment are the
non-negative integers $\{\ell_\gam\}$ represented by the vector,
\beq[eq:ell]
\ell=[\ell_1 \cdots \ell_{\nconf}]^T
\eeq
Let $\ptrue_{\alf\gam}$ denote the true probability of obtaining
outcome $\alf$ when the system is in configuration $\gam$ with state
input $\rhotrue$. Thus,
\beq[eq:av nalfgam]
\avg\ n_{\alf\gam} = \ell_\gam \ptrue_{\alf\gam}
\eeq
where the expectation $\avg(\cdot)$ taken with repect to the
underlying quantum probability distributions.  

We pose the following {\em model} of the system,
\beq[eq:palfgam]
p_{\alf\gam}(\rho) = \trace\ M_{\alf\gam} \sig_\gam(\rho)
\eeq
where $p_{\alf\gam}(\rho)$ is the outcome probability of measuring
$\alf$ when the system is in configuration $\gam$ with input state
$\rho$ belonging to the set of density matrices,
\beq[eq:rhoset]
\set{\rho\in\Cbfnn}
{\rho \geq 0,\; \trace\ \rho =1}
\eeq
$\seq{M_{\alf\gam}}$ are the POVM elements of the measurement
apparatus, and thus, for $\gam=1,\ldots,\nconf$,
\beq[eq:povm]
\sumalf\ M_{\alf\gam}=I_n,
\;\;
M_{\alf\gam} \geq 0,\
\alf=1,\ldots,\nout
\eeq
and $\sig_\gam(\rho)$ is the reduced density output state of the
$\qsys$-system model.  A general (model) representation of the $\qsys$
system is the {\em Kraus operator-sum-representation} (OSR) which can
account for many forms of error sources as well as decoherence
\cite{NielsenC:00}. Specifically, in configuration $\gam$, the
$\qsys$-system model can be parametrized by the set of {\em Kraus
matrices}, $K_\gam=\set{K_{\gam k}\in\Cbfnn}{k=1,\ldots,\kapgam}$ as
follows:
\beq[eq:osr]
\sig_\gam(\rho)
=\qsys(\rho,K_\gam)
=
\sum_{k=1}^\kapgam\ K_{\gam k} \rho K_{\gam k}^*,
\;\;\;
\sum_{k=1}^\kapgam\ K_{\gam k}^* K_{\gam k} = I_n
\eeq
with $\kapgam \leq n^2$.  Implicit in this OSR is the assumption that
the $\qsys$-system is trace preserving. Combining this with the
measurement model \refeq{povm} gives the model probability outcomes,
\beq[eq:pout osr]
p_{\alf\gam}(\rho)
=
\trace\ O_{\alf\gam}\rho,
\;\;
O_{\alf\gam} 
=
\sum_{k=1}^\kapgam\ K_{\gam k}^* M_{\alf\gam} K_{\gam k}
\eeq
In this model, the outcome probabilities are {\em linear} in the input
state density matrix. Moreover, the set
$O_\gam=\set{O_{\alf\gam}}{\alf=1,\ldots,\nout}$, satisfies
\refeq{povm}, and hence, is a POVM. \footnote{
In a more general OSR the $\qsys$-system need not be trace preserving,
hence the Kraus matrices in \refeq{osr} need not sum to identity as
shown, but rather, their sum is bounded by identity. Then the set
$O_\gam$ is not a POVM, however, satisfies,
$
\sumalf\ O_{\alf\gam} \leq I_n,
\;\;
O_{\alf\gam} \geq 0,\
\alf=1,\ldots,\nout
$
} 
If the $\qsys$-system is modeled as a unitary system, then,
\beq[eq:povm u]
\sig_\gam(\rho) = U_\gam \rho U_\gam^*,
\;\;
U_\gam^* U_\gam = I_n
\Longrightarrow
O_{\alf\gam}
=
U_\gam^* M_{\alf\gam} U_\gam
\eeq
The set $O_\gam$ is still a POVM; in effect the OSR has a single
element, namely, $K_\gam=U_\gam$.

\subsubsection*{System in the model set}

We make the following assumption throughout: {\em the true system is
in the model set}. This means that,
\beq[eq:s in m]
\ptrue_{\alf\gam} = p_{\alf\gam}(\rhotrue)
=\trace\ O_{\alf\gam}\rhotrue
\eeq
This is always a questionable assumption and in most engineering
practice is never true. Relaxing this assumption is an active research
topic particularly when identification (state or process) is to be
used for control design, \eg, see \cite{Kosut:01} and
\cite{SafonovT:97}. The case when the system is {\em not} in the model
set will not be explored any further here except for the effect of
measurement noise which is discussed next.  It is important to
emphasize that in order to produce an accurate unbiased estimate of
the true density it is necessary to know the noise elements (as
described next) which is a consequence of assumption \refeq{s in m}.

\subsubsection*{Noisy measurements}
\label{sec:noisy meas}

Sensor noise can engender more noisy outcomes than noise-free
outcomes.  Consider, for example, a photon detection device with two
photon-counting detectors.  If both are noise-free, meaning, perfect
efficiency and no dark count probability, then, provided one photon is
always present at the input of the device, there are only two possible
outcomes: $\seq{10,\ 01}$. If, however, each detector is noisy, then
either or both detectors can misfire or fire even with a photon always
present at the input. Thus in the noisy case there are {\em four}
possible outcomes: $\seq{10,\ 01,\ 11,\ 00}$. 

Let $\set{M_{\alf\gam}}{\alf=1,\ldots,\nout}$ denote the noisy POVM and
let $\set{\Mb_{\alf\gam}}{\alf=1,\ldots,\noutb}$ denote the noise-free
POVM with $\nout \geq \noutb$ where,
\beq[eq:oalfmn]
M_{\alf\gam} = \sum_{\bet=1}^\noutb \ 
\nu_{\alf\bet\gam}\
\Mb_{\bet\gam},\
\alf=1,\ldots,\nout,\
\gam=1,\ldots,\nconf
\eeq
The $\seq{ \nu_{\alf\bet\gam} }$ represents the noise in the
measurement, specifically, the conditional probability that $\alf$ is
measured given the noise-free outcome $\bet$ with the system in
configuration $\gam$. Since $\sumalf\
\nu_{\alf\bet\gam}=1,\ \forall \bet,\gam$, it follows that if the
noise-free set is a POVM then so is the noisy set.

\subsection{Maximum likelihood state estimation}
\label{sec:mle}

The Maximum Likelihood (ML) approach to quantum state estimation
presented in this section, as well as observing that the estimation is
convex, can be found in \cite{ParisETAL:01}, \cite{VDM:01} and the
references therein. Using convex programming methods, such as an
interior-point algorithm for computation, was not exploited in these
references.

If the experiments are independent, then the probability of obtaining
the data \refeq{data} is a product of the individual model
probabilities \refeq{palfgam}.  Consequently, for an {\em assumed}
initial state $\rho$, the model predicts that the probability of
obtaining the data set \refeq{data} is given by,
\beq[eq:proballdata]
\prob{D,\rho}
=
\prod_{\alf,\gam}
p_{\alf\gam}(\rho)^{n_\alf\gam}
\eeq
The data is thus captured in the outcome counts $\seq{n_{\alf\gam}}$
whereas the model terms have a $\rho$-dependence. The function
$\prob{D,\rho}$ is called the {\em likelihood} function and since it
is positive, the {\em maximum likelihood estimate} (MLE) of $\rho$ is
obtained by finding a $\rho$ in the set \refeq{rhoset} which maximizes
the {\em log-likelihood function}, or equivalently, minimizes the {\em
negative log-likelihood function},
\beq[eq:loglike1]
\bea{rcl}
L(D,\rho)
&=&
-\log\ \prob{D,\rho}
\\
&=&
\ds
-\sum_{\alf,\gam}
n_{\alf\gam}
\log\ p_{\alf\gam}(\rho)
\\
&=&
\ds
-\sum_{\alf,\gam}
n_{\alf\gam}
\log\trace\ O_{\alf\gam}\rho
\eea
\eeq
These expressions are obtained by combining \refeq{proballdata},
\refeq{loglike1} and \refeq{pout osr}.  The Maximum Likelihood state
estimate, $\rhoml$, is obtained as the solution to the optimization
problem:
\beq[eq:maxlike1]
\bea{ll}
\mbox{minimize} 
& 
L(D,\rho)
=
-\sum_{\alf,\gam}
n_{\alf\gam}
\log\trace\ O_{\alf\gam}\rho
\\
\mbox{subject to} 
& 
\rho\geq 0,\; \trace\ \rho=1
\eea
\eeq
$L(D,\rho)$ is a positively weighted sum of log-convex functions of
$\rho$, and hence, is a log-convex function of $\rho$. The constraint
that $\rho$ is a density matrix forms a convex set in $\rho$.  Hence,
\refeq{maxlike1} is in a category of a class of well studied
log-convex optimization problems, \eg, \cite{BoydV:04}.

\subsubsection*{Pure state estimation}
\label{eq:pure est}

Suppose it is known the the true state is pure, that is, $\rhotrue =
\psitrue \psitrue^*$ with $\psitrue\in\Cbf^n$ and
$\psitrue^*\psitrue=1$.  In practice we have found that solving
\refeq{maxlike1} when the true state is pure gives solutions which are
easily approximated by pure states, that is, the estimated state has
one singular value near one and all the rest are very small and
positive. 

To deal directly with pure state estimation we first need to
characterize the set of all density matrices which are pure. This is
given by the set $\set{\rho\in\Cbfnn} {\rho \geq 0,\; \rank\ \rho
=1}$, which is equivalent to,
\beq[eq:rho pure]
\set{\rho\in\Cbfnn}
{\rho \geq 0,\; \trace\ \rho =1,\; \trace\ \rho^2=1}
\eeq
The corresponding ML estimate is then the solution of,
\beq[eq:maxlike pure]
\bea{ll}
\mbox{minimize} 
& 
L(\rho)=
-\sum_{\alf,\gam}
n_{\alf\gam}
\log\trace\ O_{\alf\gam}\rho
\\
\mbox{subject to} 
& 
\rho\geq 0,\; \trace\ \rho=1,\; \trace\ \rho^2=1
\eea
\eeq
This is not a convex optimization problem because the equality
constraint, $\trace\ \rho^2=1$, is not convex.  However, relaxing this
constraint to the convex {\em inequality} constraint, $\trace\ \rho^2
\leq 1$, results in the convex optimization problem:
\beq[eq:maxlike pure rlx]
\bea{ll}
\mbox{minimize} 
& 
L(\rho)=
-\sum_{\alf,\gam}
n_{\alf\gam}
\log\trace\ O_{\alf\gam}\rho
\\
\mbox{subject to} 
& 
\rho\geq 0,\; \trace\ \rho=1,\; \trace\ \rho^2 \leq 1
\eea
\eeq
If the solution is on the boundary of the set $\trace\ \rho^2 \leq 1$,
then a pure state has been found. There is however, no guaranty that
this will occur.

\subsubsection*{Least-squares (LS) state estimation}
\label{sec:lsapp}

In a typical application the number of trials per configuration,
$\ell_\gam$, is sufficiently large so that the {\em empirical
estimate} of the outcome probability,
\beq[eq:pemp]
\pemp_{\alf\gam} = \frac{n_{\alf\gam}}{\ell_\gam}
\eeq
is a good estimate of the true outcome probability
$\ptrue_{\alf\gam}$.  The empirical probability estimate also
provides the smallest possible value of the negative log-likelihood
function, that is, $\pemp_{\alf\gam}$ is the solution to,
\beq[eq:pemp opt]
\bea{ll}
\mbox{minimize}
&
L(p)=
-\sum_{\alf,\gam}\ 
n_{\alf\gam} \log p_{\alf\gam}
\\
\mbox{subject to}
&
\sumalf\ p_{\alf\gam}=1,\forall \gam,
\;
p_{\alf\gam}\geq 0,\forall\alf,\gam
\eea
\eeq
with optimization variables $p_{\alf\gam},\forall\alf,\gam$. Thus,
for any value of $\rho$ we have the lower bound,
\beq[eq:pemp lb]
-\sum_{\alf,\gam}\ 
n_{\alf\gam} \log \frac{n_{\alf\gam}}{\ell_\gam}
\leq
-\sum_{\alf,\gam}\ 
n_{\alf\gam} \log \trace\ O_{\alf\gam} \rho
\eeq
In particular, assuming \refeq{av nalfgam} holds, and the $\ell_\gam$
trials are independent, then the variance of the empirical estimate is
known to be \cite{Papoulis:65},
\beq[eq:var pemp]
\var\ \pemp_{\alf\gam}
=
\frac{1}{\ell_\gam}
\ptrue_{\alf\gam}
\left( 1-\ptrue_{\alf\gam} \right)
\eeq
It therefore follows that for large $\ell_\gam$, $\pemp_{\alf\gam}
\approx \ptrue_{\alf\gam}$, and if as assumed \refeq{s in m}, the
system is in the model set, then $\ptrue_{\alf\gam} = \trace\
O_{\alf\gam} \rhotrue$.  These two conditions lead to taking the state
estimate as the solution to the constrained weighted least-squares
problem:
\beq[eq:lsopt cvx]
\bea{ll}
\mbox{minimize}
&
\sum_{\alf,\gam}\ 
w_\gam
\left( \pemp_{\alf\gam} - \trace\ O_{\alf\gam} \rho \right)^2
\\
\mbox{subject to}
&
\rho\geq 0,\; \trace\ \rho=1
\eea
\eeq
The weights, $w_\gam$, are chosen by the user to emphasize different
configurations. A typical choice is the distribution of experiments
per configuration, hence, $w_\gam\geq 0,\ \sum_\gam\
w_\gam=1$. Because of the semi-definite constraint, this
weighted-least-squares problem is a convex optimization in the
variable $\rho$. For large $\ell_\gam$, the solution ought to be a
good estimate of the true state. There is, however, little numerical
benefit in solving \refeq{lsopt cvx} as compared to \refeq{maxlike1}
-- they are both convex optimization problems and the numerical
complexity is similar provided \refeq{maxlike1} is solved using an
interior-point method \cite{BoydV:04}. Some advantage is obtained by
dropping the semidefinite constraint $\rho\geq 0$ in \refeq{lsopt cvx}
resulting in,
\beq[eq:lsopt]
\bea{ll}
\mbox{minimize}
&
\sum_{\alf,\gam}\ 
w_\gam
\left( \pemp_{\alf\gam} - \trace\ O_{\alf\gam} \rho \right)^2
\\
\mbox{subject to}
&
\trace\ \rho=1
\eea
\eeq
This is a standard least-squares problem with a linear equality
constraint which can be solved very efficiently using a singular value
decomposition to eliminate the equality constraint
\cite{GolubV:83}. For sufficiently large $\ell_\gam$ the resulting
estimate {\em may} satisfy the positivity constraint $\rho\geq 0$. If
not, it is usually the case that some of the small eigenvalues of the
state estimate or estimated outcome probabilities are slightly
negative which can be manually set to zero. Solving \refeq{lsopt} {\em
is} numerically faster than solving \refeq{maxlike1}, but not by
much. Even with a large amount of data the solution to \refeq{lsopt}
can produce estimates which are not positive if the data is not
sufficiently rich. In this case the estimates from any procedure which
eliminates the positivity constraint can be very misleading.

It thus appears that even for large $\ell_\gam$, there is no
significant benefit accrued, either because of numerical precision or
speed, to using the empirical estimate followed by standard
least-squares. If, however, the $\ell_\gam$ are not sufficiently large
and/or the data is not sufficiently rich, then it is unlikely that the
estimate from \refeq{lsopt} will be accurate.

One possible advantage does come about because the solution to
\refeq{lsopt} can be expressed analytically, and thus it is possible
to gain an understanding of how to select the POVM. For example, in
\cite{NielsenC:00} special POVM elements are selected to essentially
diagonalize the problem, thereby making the least-squares problem
\refeq{lsopt} simpler, \ie, the elements of the density matrix can be
estimated one at a time. However, implementing the requisite POVM set
may be very difficult depending on the physical apparatus involved.

\subsection{Experiment design for state estimation}
\label{sec:expt des}

In this section we describe the experiment design problem for quantum
state estimation. The objective is to select the number of experiments
per configuration, the elements of the vector $\ell=[\ell_1 \cdots
\ell_\nconf]^T\in\Rbf^\nconf$, so as to minimize the error between the
state estimate, $\rhoh(\ell)$, and the true state
$\rhotrue$. Specifically, we would like to solve for $\ell$ from:
\beq[eq:expdes 0]
\bea{ll}
\mbox{minimize}
&
\avg\ \norm{\rhoh(\ell)-\rhotrue}_{\rm frob}^2
\\
\mbox{subject to}
&
\sumgam\ \ell_\gam = \nex
\\
&
\mbox{integer}\ \ell_\gam \geq 0,\;
\gam=1,\ldots,\nconf
\eea
\eeq
where $\nex$ is the desired number of total experiments.  This is a
difficult, if not insoluble problem for several reasons. First, the
solution depends on the estimation method which produces
$\rhoh(\ell)$. Secondly, the problem is integer combinatorial because
$\ell$ is a vector of integers. And finally, the solution depends on
$\rhotrue$, the very state to be estimated. Fortunately all these
issues can be circumvented.

We first eliminate the dependence on the estimation method. The
following result can be established using the {\em \crao Inequality}
\cite{Cramer:46}. The derivation is in Appendix \refsec{var rho}.

\bquote
{\bf State estimation variance lower bound} \footnote{
\crao bounds previously reported in the literature are not quite
correct as they do not include the linear constraint $\trace\ \rho=1$
as is done here.
%
}
\\
Suppose the system generating the data is in the model set used for
estimation, \ie, \refeq{s in m} holds.  For $\ell=[\ell_1 \cdots
\ell_{\nconf}]$ experiments per configuration, suppose $\rhoh(\ell)$
is a density matrix {\em and} an unbiased estimate of $\rhotrue$, \ie,
$\rhoh(\ell) \geq 0$, $\trace\ \rhoh(\ell)=1$, and $\avg\ \rhoh(\ell)
= \rhotrue$. Under these conditions, the estimation error variance
satisfies,
\beq[eq:var rho]
\avg\ \norm{\rhoh(\ell)-\rhotrue}_{\rm frob}^2
\geq
V(\ell,\rhotrue)
=
\trace\ G(\ell,\rhotrue)^{-1}
\eeq
where\footnote{
The $\vec$ operation takes the rows of a matrix and stacks them one
row at a time on top of each other. Two useful expressions are
$\vec(AXB)=(B^T\otimes A)\vec\ X$ and $\trace\ AX=(\vec\ A^T)^T \vec\
X$.
}
\beq[eq:var rho 1]
\bea{rcl}
\ds
G(\ell,\rhotrue) 
&=& 
\ds
\sum_{gam=1}^\nconf\ \ell_\gam G_\gam(\rhotrue)
\in\Rbf^{n^2-1 \times n^2-1}
\\&&\\
\ds
G_\gam(\rhotrue)
&=&
\ds
\Ceq^T
\left(
\sumalf\
\frac{a_{\alf\gam} a_{\alf\gam}^*}{p_{\alf\gam}(\rhotrue)}
\right)
\Ceq
\in\Rbf^{n^2 \times n^2}
\\&&\\
a_{\alf\gam} &=& \vec\ O_{\alf\gam} \in \Cbf^{n^2}
\eea
\eeq
and $\Ceq\in\Rbf^{n^2 \times n^2-1}$ is part of the unitary matrix in
the singular value decomposition:
$ \vec\ I_n = USW^T\in\Rbf^{n^2},\ W=[c\ \Ceq]\in\Rbf^{n^2\times n^2}$.
\equote
This theorem states the for any unbiased estimate of $\rhotrue$, the
variance of the estimate satisfies the inequality \refeq{var rho}. The
power of the result is that it is independent of {\em how} the
estimate is obtained, \ie, no estimation algorithm which produces an
unbiased estimate can have an estimation error variance smaller than
that in \refeq{var rho}. There is a generalization for biased
estimators but we will not pursue that here.

In general it is difficult to determine if any estimate will achieve
the lower bound. However, under the conditions stated in the above
result, the ML estimate, $\rhoml(\ell)$, the solution to
\refeq{maxlike1}, approaches $\rhotrue$ with probability one,
asymptotically as $\nex$ increases, and the asymptotic distribution
becomes Gaussian with covariance given by the \crao bound (see
\refsec{crao} for the covariance expression and \cite{Ljung:87} for a
derivation).

The one qualifier to the \crao bound as presented is that the
indicated inverse exists. This condition, however, is necessary and
sufficient to insure that the state is identifiable. More precisely,
the state is {\em identifiable} if and only if,
\beq[eq:identifiable]
G(\ell=1_\nconf,\ \rhotrue)
=
\Ceq^*
\left(
\sumgam\sumalf\
\frac{a_{\alf\gam} a_{\alf\gam}^*}{p_{\alf\gam}(\rhotrue)}
\right)
\Ceq
\;\;\;
\mbox{is invertible}
\eeq
Under the condition of identifiability, the experiment design problem
can be expressed by the following optimization problem in the vector
of integers $\ell$:
\beq[eq:expdes]
\bea{ll}
\mbox{minimize}
&
V(\ell,\rhotrue) = \trace\ G(\ell,\rhotrue)^{-1}
\\
\mbox{subject to}
&
\sumgam\ \ell_\gam = \nex
\\
&
\mbox{integer}\ \ell_\gam \geq 0,\;
\gam=1,\ldots,\nconf
\eea
\eeq
where $\nex$ is the desired number of total experiments.  The good
news is that the objective, $V(\ell,\rhotrue)$, is convex in $\ell$
\cite[\S 7.5]{BoydV:04}. Unfortunately, there are still two
impediments: (i) restricting $\ell$ to a vector of integers makes the
problem combinatorial; (ii) the lower-bound function
$V(\ell,\rhotrue)$ depends on the true value, $\rhotrue$.  These
difficulties can be alleviated to some extent. For (i) we can use the
convex relaxation described in \cite[\S 7,5]{BoydV:04}. For (ii) we
can solve the relaxed experiment design problem with either a set of
``what-if'' estimates as surrogates for $\rhotrue$, or use nominal
values to start and then ``bootstrap'' to more precise values by
iterating between state estimation and experiment design. We now
explain how to perform these steps.

\subsubsection*{Relaxed experiment design for state estimation}

Following the procedure in \cite[\S 7.5]{BoydV:04}, introduce the
variables $\lam_\gam=\ell_\gam/\nex$, each of which is the fraction of
the total number of experiments performed in configuration
$\gam$. Since all the $\ell_\gam$ and $\nex$ are non-negative
integers, each $\lam_\gam$ is non-negative and {\em rational},
specifically an integer multiple of $1/\nex$, and in addition,
$\sum_\gam \lam_\gam=1$.  Let $\rhosurr$ denote a surrogate for
$\rhotrue$, \eg, an estimate or candidate value of $\rhotrue$. Using
\refeq{var rho}-\refeq{var rho 1} gives,
\beq[eq:nexscale]
V(\ell=\nex\lam,\rhosurr) = \frac{1}{\nex}V(\lam,\rhosurr)
\eeq
Using \refeq{var rho 1},
\beq[eq:vlamrho]
\bea{rcl}
V(\lam,\rhosurr)
&=&
\trace\
G(\lam,\rhosurr)^{-1}
\\
\ds
G(\lam,\rhosurr) 
&=& 
\ds
\sumgam\ \lam_\gam G_\gam(\rhosurr)
\eea
\eeq
Hence, the objective function $V(\ell,\rhosurr)$ can be replaced with
$V(\lam,\rhosurr)$ and the experiment design problem \refeq{expdes} is
equivalent to.
\beq[eq:expdes rat]
\bea{ll}
\mbox{minimize}
&
V(\lam,\rhosurr) = \trace\ G(\lam,\rhosurr)^{-1}
\\
\mbox{subject to}
&
\sumgam\ \lam_\gam = 1
\\
&
\lam_\gam \geq 0,\;
\mbox{integer multiple of $1/\nex$},\;
\gam=1,\ldots,\nconf
\eea
\eeq
The objective is now a convex function of the $\lam_\gam$, but it is
still a combinatorial problem because the $\lam_\gam$ are constrained
to each be an integer multiple of $1/\nex$.  If $\lam_\gam$ is only
otherwise constrained to the non-negative reals, then this has the
effect of relaxing the constraint that the $\ell_\gam$ are integers.
As phrased in \cite{BoydV:04}, the {\em relaxed} experiment design
problem is:
\beq[eq:rel expdes]
\bea{ll}
\mbox{minimize}
&
V(\lam,\rhosurr)
=
\trace\left(
\sumgam\ \lam_\gam G_\gam(\rhosurr)
\right)^{-1}
\\
\mbox{subject to}
&
\sumgam\ \lam_\gam = 1
\\
&
\lam_\gam\geq 0,\;
\gam=1,\ldots,\nconf
\eea
\eeq
The objective is convex, the equality constraint is linear, and the
inequality constrains are convex, hence, this is a convex optimization
problem in $\lam\in\Rbf^\nconf$.  Let $\lamopt$ denote the optimal
solution to \refeq{rel expdes}. Since the problem no longer depends on
$\nex$, $\lamopt$ can be viewed as a distribution of experiments per
configuration.\footnote{
{\bf Caveat emptor}: The relaxed optimal experiment design
distribution, $\lamopt$, is optimal with respect to the initial state
$\rhosurr$, a surrogate for $\rhotrue$. Thus, $\lamopt$ is {\em not}
optimal with respect to $\rhotrue$. This should be no surprise because
the underlying goal is to find a good estimate of $\rhotrue$.
}
Clearly there is no guaranty that $\nex\lamopt$ is a vector of integer
multiples of $1/\nex$. A practical choice for obtaining a vector of
integer multiples of $1/\nex$ is,
\beq[eq:ellrnd]
\ellrnd = \round{\nex \lamopt}
\eeq
If $\ell^\opt$ is the (unknown) integer vector solution to \refeq{expdes},
then we have the relations:
\beq[eq:bnds]
V(\ellrnd,\rhosurr)
\geq 
V(\ell^\opt,\rhosurr)
\geq
V(\nex\lamopt,\rhosurr) 
\eeq
The optimal objective is thus bounded above and below by known values
obtained from the relaxed optimization.  The gap within which falls
the optimal solution can be no worse than the difference between
$V(\ellrnd,\rhosurr)$ and $V(\nex\lamopt,\rhosurr)$, which can be computed
solely from $\lamopt$. If the gap is sufficiently small then for all
practical purposes the ``optimal'' solution is $\lamopt$. From now on
we will refer to $\lamopt$ as the optimal solution rather than the
relaxed optimal.

\subsubsection*{Performance tradeoff}

The optimal distribution $\lamopt$ can be used to guide the
elimination of small values of $\lamopt$. For example, consider the
{\em suboptimal} distribution, $\lamsub$, obtained by selecting the
largest $\nr$ out of $\nconf$ non-zero values of $\lamopt$. Let
$\nexsub$ denote the integer vector of configurations,
\beq[eq:nexsub]
\nexsub=\round{\nex\lamsub}
\eeq
Using \refeq{nexscale}, the minimum number of experiments so that
$V(\nexsub,\rhosurr)\leq V_0$ is given by,
\beq[eq:nexmin]
\nexsub =\round{V(\lamsub,\rhosurr)/V_0}
\eeq
As $\nr$ is varied, the graph $\{\nr,\ \nexsub\}$ establishes a
tradeoff between the number of configurations per experiment versus
the total number of experiments such that the lower bound on the
estimation variance does not exceed the desired value $V_0$. When
$\nr=\nconf$, \refeq{nexmin} is identical with \refeq{ellrnd}.

The condition number of the matrix $G(\lamopt,\rhosurr)$ gives an
indication of the identifiability of the density matrix $\rhosurr$. A
very large condition number means that the linear combination of
elements of the density matrix associated with the small eigenvalue
will be more difficult to obtain then those combinations associated
with a large eigenvalue. The condition number of $G(\lamopt,\rhosurr)$ is
not only affected by the number of experiments per configuration,
$\lamopt$, but by the configurations themselves. Examining
$G(\lamopt,\rhosurr)$ for different $\rhosurr$ (surrogates of $\rhotrue$)
and different configurations can help establish a good experiment
design.

\subsubsection*{Bootstrapping}

A standard approach used to circumvent not knowing the true state
needed to optimize the experiment design is to proceed adaptively, or
by ``bootstrapping.'' The idea is to use the current estimate of the
initial state found from \refeq{maxlike1}, then solve \refeq{rel
expdes}, and then repeat. The algorithm at the $k$-th iteration looks
like this:
\beq 
\bea{rcl}
\rhoh(k)
&=&
\ds
\arg\min_{\rho}
V(\ellh(k-1),\rho)
\\&&\\
\lamopt(k) 
&=&
\ds
\arg\min_{\lam} 
V(\lam,\rho=\rhoh(k))
\\&&\\ 
\ellh(k) 
&=&
\ds 
\round{\nex \lamopt(k)}
\eea 
\eeq 
The initial distribution $\ellh(0)$ could be chosen as uniform, \eg,
the same for a not too large number of configurations. The algorithm
could also start by first solving for a distribution from an initial
state surrogate. In each iteration we could also vary $\nex$. Although
each optimization is convex, the joint problem may not be. Conditions
for convergence would need to be investigated as well as establishing
that this method is efficient, \ie, reduces the number of trials.  We
will not pursue this any further here.

\subsubsection*{Dual experiment design problem}
\label{sec:dual}

{\em Lagrange Duality Theory} can provide a lower bound on the
objective function in an optimization problem as well as establishing
optimality conditions often leading to insights into the optimal
solution structure \cite[Ch.5]{BoydV:04}. In many cases the largest
lower bound -- the solution of the {\em dual problem} -- is equal to
the optimal objective function. The dual problem associated with the
experiment design problem \refeq{rel expdes} is,
\beq[eq:dual expdes]
\bea{ll}
\mbox{maximize}
&
\left( \trace\ W^{1/2} \right)^2
\\
\mbox{subject to}
&
\trace\ W G_\gam(\rhosurr) \leq 1,
\;
\gam=1,\ldots,\nconf
\\
&
W > 0
\eea
\eeq
The optimization variable is $W\in\Cbf^{n^2-1 \times n^2-1}$. The
above form of the dual is given in \cite[\S 7.5.2]{BoydV:04} for a
slightly simpler problem (the $G_\gam$ are dyads) but is essentially
the same. A key observation arises from the {\em complementary
slackness condition},
\beq[eq:cs cond]
\lamopt_\gam
\left(
\trace\ W^\opt G_\gam(\rhosurr) - 1
\right)
=0,
\;\;
\gam=1,\ldots,\nconf
\eeq
where $\lamopt$ is the solution to the {\em primal problem},
\refeq{rel expdes}, and $W^\opt$ is the solution to the dual problem,
\refeq{dual expdes}.  Thus, only when the equality constraint holds,
$\trace\ W^\opt G_\gam(\rhosurr)=1$, is the associated $\lamopt_\gam$ not
necessarily equal to zero. It will therefore be usually the case that
many of the elements of the optimal distribution will be zero.

Strong duality also holds for this problem, thus the optimal primal
and dual objective values are equal,
\beq[eq:strdual]
\trace\left(
\sumgam\ \lamopt_\gam G_\gam(\rhosurr)
\right)^{-1}
=
\left( \trace\ (W^\opt)^{1/2} \right)^2
\eeq
For this problem, a pair $(\lam,\ W)$ is optimal with respect to
$\rhosurr$ if and only if:
\beq[eq:oc expdes]
\bea{rcl}
\sumgam\ \lam_\gam &=& 1
\\
\lam_\gam &\geq& 0,\ \forall\gam
\\
\lamopt_\gam
\left(
\trace\ W^\opt G_\gam(\rhosurr) - 1
\right)
&=& 0,\ \forall\gam
\\
\trace\ W G_\gam(\rhosurr) &\leq& 1,\ \forall\gam
\\
\trace\left(
\sumgam\ \lam_\gam G_\gam(\rhosurr)
\right)^{-1}
&=&
\left( \trace\ (W^\opt)^{1/2} \right)^2
\eea
\eeq
%


\subsection{Example: experiment design for state estimation}
\label{sec:example qs}

\begin{figure}[h]
\centering
\epsfxsize=4in
\epsfboxit{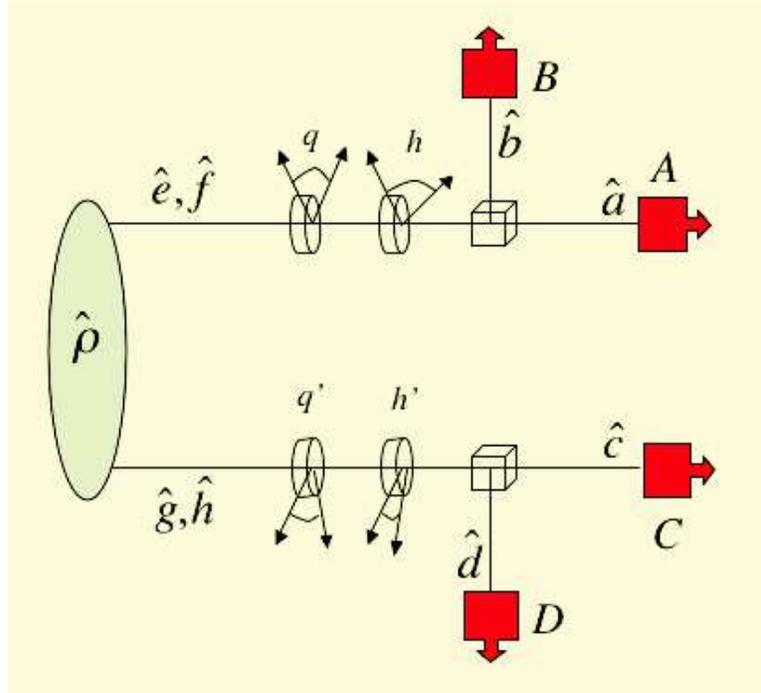}
\caption{Detection apparatus for two-photon tomography}
\label{fig:tomag01}
\end{figure}

A schematic of an apparatus for state tomography of a pair of
entangled photons specified by the quantum state (density matrix)
$\rho$ is shown in figure \reffig{tomag01}.

The set up has four photon-counting detectors, A, B, C, D. There are
four continuous variable settings for the quarter-wave plates and
half-wave plates, \ie, $q,h,q',h'$. For any settings of these
parameters one of the detectors in each arm will register a
photon. The objective is to determine the optimal settings of these
parameters and the number of experiments per setting for estimation of
the state $\rho$ of the pair using as data the photon counts from the
four detectors.

Because the photon sources are not completely efficient, the input
quantum state actually consists of either two or zero photons. The
detectors register a 0 or 1 depending on whether a photon is incident
on them or not.  The basis states for the upper arm are therefore:
$\ket{0}_e\ket{0}_f,\ \ket{0}_e\ket{1}_f,\ \ket{1}_e\ket{0}_f$.  There
is a similar set for the lower arm (modes $g, h$).
 
The firing patterns for an arbitrary setting of the wave plates,
assuming perfect detection efficiency and no dark counts are given in
the table:
\bc
\btab{|c|c|c|c|}
\hline
A & B & C & D \\
\hline
0 & 1 & 0 & 1 \\
0 & 1 & 1 & 0 \\
1 & 0 & 0 & 1 \\
1 & 0 & 1 & 0 \\
0 & 0 & 0 & 0 \\
\hline
\etab
\ec
The probabilities for these patterns are given by
\beq
p_{ijk\ell}
=
\trace\ (M_{AB}^{ij} \otimes M_{CD}^{k\ell}) \rho
\eeq
where $\{i,j,k,\ell\}\in\{0,1\}$, and $M_{AB}^{ij}$ is the projector
for detector A to register count $i$ and simultaneously detector B to
register count $j$. Similarly, $M_{CD}^{k\ell}$ is the projector for
detector C to register count $j$ and simultaneously detector D to
register count $\ell$.  The projectors for A and B in the above basis
are:
\beq[eq:projmat]
\bea{rcl}
\ds
M_{AB}^{00}
&=&
\left[
\bea{ccc}
1 & 0 & 0\\
0 & 0 & 0\\
0 & 0 & 0
\eea
\right]
\\&&\\
\ds
M_{AB}^{10}
&=&
\ds
\left[
\bea{c}
0
\\
\psi_1(h,q)
\eea
\right]
\left[
0\ \psi_1(h,q)^*
\right],\
\psi_1(h,q)
=
\frac{1}{\sqrt{2}}
\left[
\bea{c}
\sin 2h+i\sin 2(h-q)
\\
\cos 2h-i\cos 2(h-q)
\eea
\right]
\\&&\\
\ds
M_{AB}^{01}
&=&
\ds
\left[
\bea{c}
0
\\
\psi_2(h,q)
\eea
\right]
\left[
0\ \psi_2(h,q)^*
\right],\
\psi_2(h,q)
=
\frac{1}{\sqrt{2}}
\left[
\bea{c}
\cos 2h+i\cos 2(h-q)
\\
-\sin 2h+i\sin 2(h-q)
\eea
\right]
\eea
\eeq
A similar set of projectors can be written for C and D with the
variables $h,q$ replaced by their primed counterparts
$h',q'$.

The protocol is to measure the probabilities for enough settings of
the variables that the elements of the two-photon density operator can
be estimated. The two-photon density operator is the direct product of
the one-photon density operator, for which the set of 3 states given
above forms a basis. The basis states of the two-photon $(9\times 9)$
density operator, $\rho$, are:
$\ket{ijk\ell}=\ket{i}_e\ket{j}_f\ket{k}_g\ket{\ell}_h$
with $i,j,k,\ell \in \{0,1\}$
Hence,
\[
\bea{ll}
\mbox{zero photon} & \ket{0000} \\
\mbox{one photon} & \ket{0100},\ \ket{1000},\ \ket{0001},\ \ket{0010} \\
\mbox{two photon} & \ket{0101},\ \ket{0110},\ \ket{1001},\  \ket{1010}
\eea
\]
%

\subsubsection*{simulation results: one-arm}

Consider only one arm of the apparatus in figure \reffig{tomag01}, say
the upper arm with detectors (A,B).  Suppose the wave plate settings
are,
\beq
\set{h_\gam,\ q_\gam}{\gam=1,\ldots,\nconf}
\eeq 
Assume also that the incoming state {\em always} is one photon, never
none. Hence, $\rho\in\Cbf^{2\times 2}$ and the projectors are:
\beq[eq:povm hqgam]
\bea{rcl}
\ds
M_\gam^{10}
&=&
\psi_1(h_\gam,q_\gam)
\psi_1(h_\gam,q_\gam)^*
\\&&\\
\ds
M_\gam^{01}
&=&
\psi_2(h_\gam,q_\gam)
\psi_2(h_\gam,q_\gam)^*
\eea
\eeq
with $\psi_1,\psi_2$ from \refeq{projmat}.  Assuming each detector has
efficiency $\eta,\ 0\leq\eta\leq 1$ and a non-zero dark count
probability, $\del,\ 0\leq\del\leq 1$, then there are four possible
outcomes at detectors A,B given in the following table:
\bc
\btab{|c||c|c|}
\hline
$\alf$ & A & B \\
\hline
10 & 1 & 0 \\
01 & 0 & 1 \\
00 & 0 & 0 \\
11 & 1 & 1 \\
\hline
\etab
\ec
Following \cite{GriceW:96,WalW:98} the probability of a dark count is
denoted by the conditional probability,
\beq
\pcyn=\del
\eeq
where $1|0$ means the detector has fired ``1'' given that no photon is present
at the detector ``0.'' As shown in \cite{GriceW:96}, it therefore follows
that the probability that the detector does not fire ``0'' although a
photon is present at the detector ``1'' is given by,
\beq
\pcny=\pny
\eeq
Here $1-\eta$ is the probability of no detection and $1-\del$ is the
probability of no dark count. The remaining conditional probabilities
are, by definition, constrained to obey:
\beq
\bea{rcl}
\pcyn + \pcnn &=& 1
\\
\pcyy + \pcny &=& 1
\eea
\eeq
The probabilities for the firing patterns in the above table are thus
given by \refeq{palfgam} with the following observables $M_{\alf\gam}$: 
\beq
\bea{rcl}
\ds
M_{10,\gam}
&=&
\ds
\pcyy\pcnn M_\gam^{10}+\pcyn \pcny M_\gam^{01}
\\&&\\
\ds
M_{01,\gam}
&=&
\ds
\pcny \pcyn M_\gam^{10}+\pcnn \pcyy M_\gam^{01}
\\&&\\
\ds
M_{00,\gam}
&=&
\ds
\pcny \pcnn M_\gam^{10}+\pcnn \pcny M_\gam^{01}
\\&&\\
\ds
M_{11,\gam}
&=&
\ds
\pcyy \pcyn M_\gam^{10}+\pcyn \pcyy M_\gam^{01}
\eea
\eeq
Numerical computer simulations were performed for two input state cases: 
\beq[eq:rhot]
\bea{ll}
\mbox{\bf pure state:}
&
\ds
\rhopur
= 
\frac{1}{2}
\left[
\bea{cc}
1 & 1\\
1 & 1
\eea
\right]
=
\psi_0 \psi_\ast,\
\psi_0
=
\frac{1}{\sqrt{2}}
\left[\bea{c} 1 \\ 1 \eea \right]
\\&\\
\mbox{\bf mixed state:}
&
\ds
\rhomix
= 
\left[
\bea{cc}
0.6 & -0.2i\\
0.2i & 0.4
\eea
\right]
\eea
\eeq
For each input state case we computed $\lamh$ with and without ``noise:''
\beq
\bea{rl}
\mbox{no noise}
&
\left\{
\bea{ll}
\mbox{detector efficiency} & \eta=1
\\
\mbox{dark count probability} & \del=0
\eea
\right.
\\&\\
\mbox{yes noise}
&
\left\{
\bea{ll}
\mbox{detector efficiency} & \eta=0.75
\\
\mbox{dark count probability} & \del=0.05
\eea
\right.
\eea
\eeq
For all cases and noise conditions we used the wave plate settings:
\beq[eq:hq1arm]
\bea{l}
h_i=(i-1)(5^\circ),\ i=1,\ldots,10
\\
q_i=(i-1)(5^\circ),\ i=1,\ldots,10
\eea
\eeq
Both angles are set from $0$ to $45^\circ$ in $5^\circ$
increments. This yields a total of $\nconf=10^2=100$ configurations
corresponding to all the wave plate combinations.  

\begin{figure}[h]
\epsfxsize=6.5in
\epsfboxit{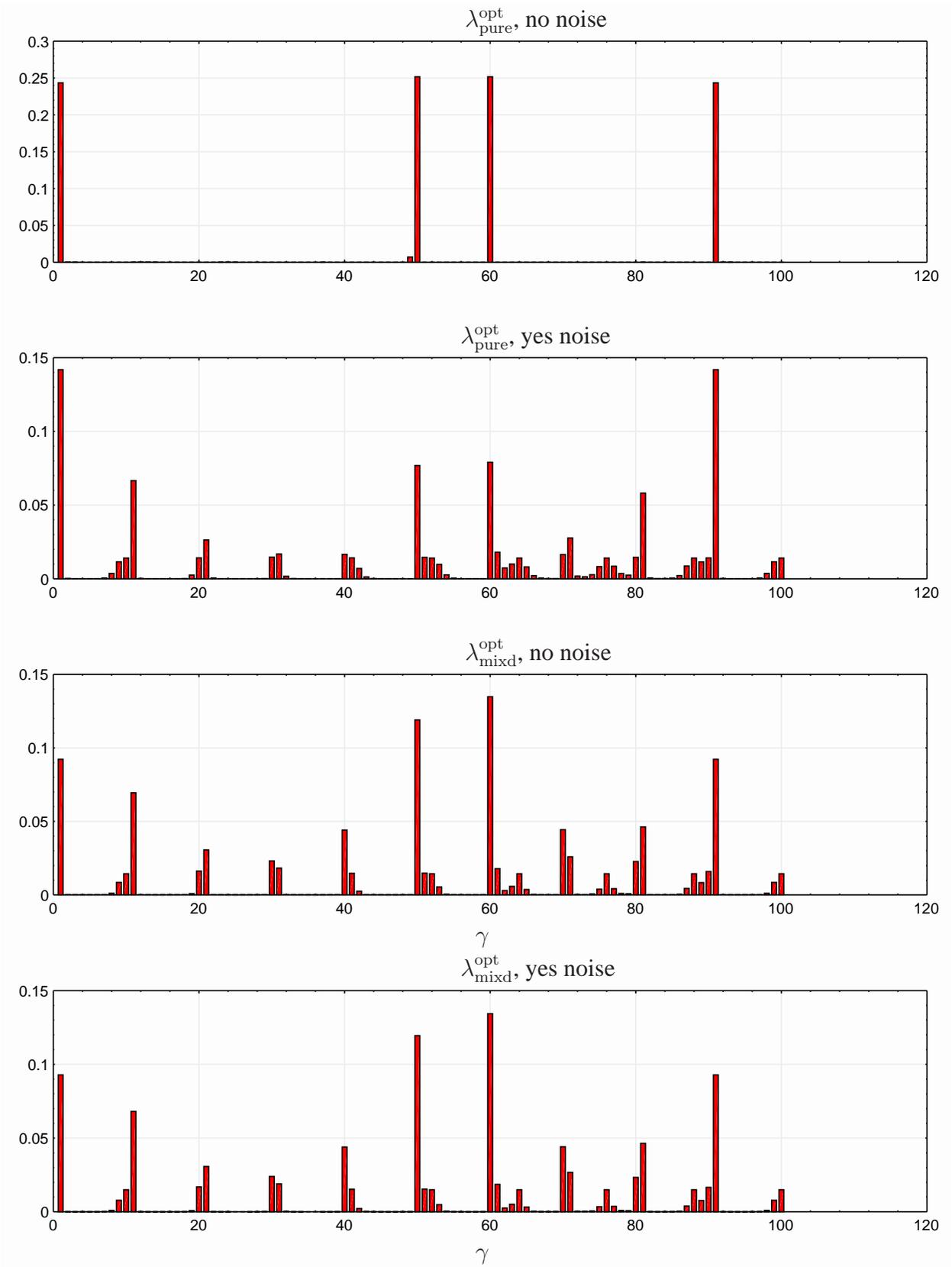}
\caption{optimal distributions -- one-arm}
\label{fig:lamopt1arm}
\end{figure}

Figure \reffig{lamopt1arm} shows the optimal distributions $\lamopt$
versus configurations $\gam=1,\ldots,100$ for all four test cases: two
input states with and without noise. Observe that the optimal
distributions are {\em not} uniform but are concentrated near
the same particular wave plate settings. These settings are very close
to those established in \cite{JamesETAL:01}.

To check the gap between the relaxed optimum $\lamh$ and the unknown
integer optimum we appeal to \refeq{ellrnd}-\refeq{bnds}. The
following table shows that these distributions are good approximation
to the unknown optimal integer solution for even not so large $\nex$
for the two state cases with no noise. Similar results were obtained
for the noisy case.
\[
\bea{|c||c|c|}
\hline
&&\\
\nex 
& 
\fracds{V(\nex{\lamopto},\rhopur)}{V({\ellrnd(\rhopur)},\rhopur)}
&
\fracds{V(\nex{\lamoptt},\rhomix)}{V({\ellrnd(\rhomix)},\rhomix)}
\\
&&\\
\hline
100 & .9797 & .7761
\\
1000 & .9950 & .9735
\\
10,000 & .9989 & .9954
\\
\hline
\eea
\]
The following table compares the distributions for optimal, suboptimal
with 8 angles, uniform at the 8 suboptimal angles, and uniform at all
100 angles by computing the minimum number of experiments required to
obtain an RMS estimation error of no more than 0.01.
\[
\renewcommand{\arraystretch}{1}
\bea{|c||c|c|c|c|}
\hline
\mbox{input state} 
& 
\bea{c}
\mbox{optimal}
\\
\round{\nex \lamopt}
\\
\nconf=100
\eea
& 
\bea{c}
\mbox{suboptimal}
\\
\round{\nex\lamsub}
\\
\nconf=8
\eea
&
\bea{c}
\mbox{uniform}
\\ 
(1/8)\ {\bf 1}(\lamsub)
\\
\nconf=8
\eea
&
\bea{c}
\mbox{uniform}
\\ 
(1/100)\ {\bf 1}(\lamopt)
\\
\nconf=100
\eea
\\
\hline
\mbox{$\rhopur$, no noise} & 20,308 & 20,308 & 20,638 & 29,274\\
\mbox{$\rhopur$, yes noise} & 37,775 & 64,129 & 40,178 & 52,825\\
\mbox{$\rhomix$, no noise} & 41,890 & 92,471 & 69,750 & 64,780\\
\mbox{$\rhomix$, yes noise} & 61,049 & 134,918 & 101,425 & 94,385\\
\hline
\eea
\]
For these optical experiments there is significant cost (in time)
associated with changing wave plate angles and very little cost (in
time) for an experiment. As a result, although the uniform
distribution at all 100 angles does not require a significant increase
in the number of experiments, and in some cases fewer experiments than
the suboptimal case, it is {\em very} costly in terms of changing wave
plate angles. The following table shows the wave plate settings and
suboptimal distributions from the above table.
\[
\bea{|c|c||c|c|c|c|}
\hline
  &  & \lamr(\rhopur) & \lamr(\rhopur) & \lamr(\rhomix) & \lamr(\rhomix) \\
h  & q   & \mbox{no noise} & \mbox{yes noise} 
& \mbox{no noise} & \mbox{yes noise}\\
\hline 
0     &    0     &    0.24    &    0.23    &    0.14    &    0.14    \\    
  5     &    0     &    0       &    0.11    &    0.11    &    0.11    \\    
  10    &    0     &    0       &    0.04    &    0       &    0       \\    
  15    &    45    &    0       &    0       &    0.07    &    0.07    \\    
  20    &    40    &    0.01    &    0       &    0       &    0       \\    
  20    &    45    &    0.25    &    0.12    &    0.19    &    0.19    \\    
  25    &    45    &    0.25    &    0.13    &    0.21    &    0.21    \\    
  30    &    45    &    0       &    0       &    0.07    &    0.07    \\    
  35    &    0     &    0       &    0.04    &    0       &    0       \\    
  40    &    0     &    0       &    0.09    &    0.07    &    0.07    \\    
  45    &    0     &    0.24    &    0.23    &    0.14    &    0.14    \\    
  45    &    5     &    0       &    0       &    0       &    0       \\    
\hline
\eea
\]
This table shows that only a few wave plate angle changes are
necessary if the suboptimal distributions are invoked. And from the
previous table, as already observed, the suboptimal settings do not
require a significant increase in the number of experiments required
to achieve the desired estimation accuracy.

\psfrag{gam}{$\gam$}
\psfrag{r01}{$\lamopt_\pur$, no noise}
\psfrag{r01n}{$\lamopt_\pur$, yes noise}
\psfrag{r02}{$\lamopt_\mix$, no noise}
\psfrag{r02n}{$\lamopt_\mix$, yes noise}

\clearpage
\subsubsection*{simulation results: both arms}

We explore the two arm case under the assumption that two photons are
always present at the input, thereby excluding the zero photon
case.\footnote{
In an actual laboratory setting it is important to include the zero
photon case; it is often that no photon is actually present at the
input.
}
The table of detector firing patterns is:
\bc
\btab{|c||c|c|c|c|}
\hline
$\alf$ & A & B & C & D \\
\hline
0101 & 0 & 1 & 0 & 1 \\
0110 & 0 & 1 & 1 & 0 \\
1001 & 1 & 0 & 0 & 1 \\
1010 & 1 & 0 & 1 & 0 \\
\hline
\etab
\ec
The following three $4\times 4$ input states are considered:
\beq[eq:rhoott]
\bea{l}
\rhopur\otimes\rhopur
\\
\rhopur\otimes\rhomix
\\
\rhomix\otimes\rhomix
\eea
\eeq
with $\rhopur,\ \rhomix$ given by \refeq{rhot}.  We use the angles
from the one-arm optimal distribution for the largest 8 values:
\beq
\bea{ll}
h\in[0,20,25,45] & q\in[0,20,25,45]
\\
h'\in[0,20,25,45] & q'\in[0,20,25,45]
\eea
\eeq
This yields a total of $\nconf=4^4=256$ wave plate settings.  

\begin{figure}[h]
\epsfxsize=6.5in
\epsfboxit{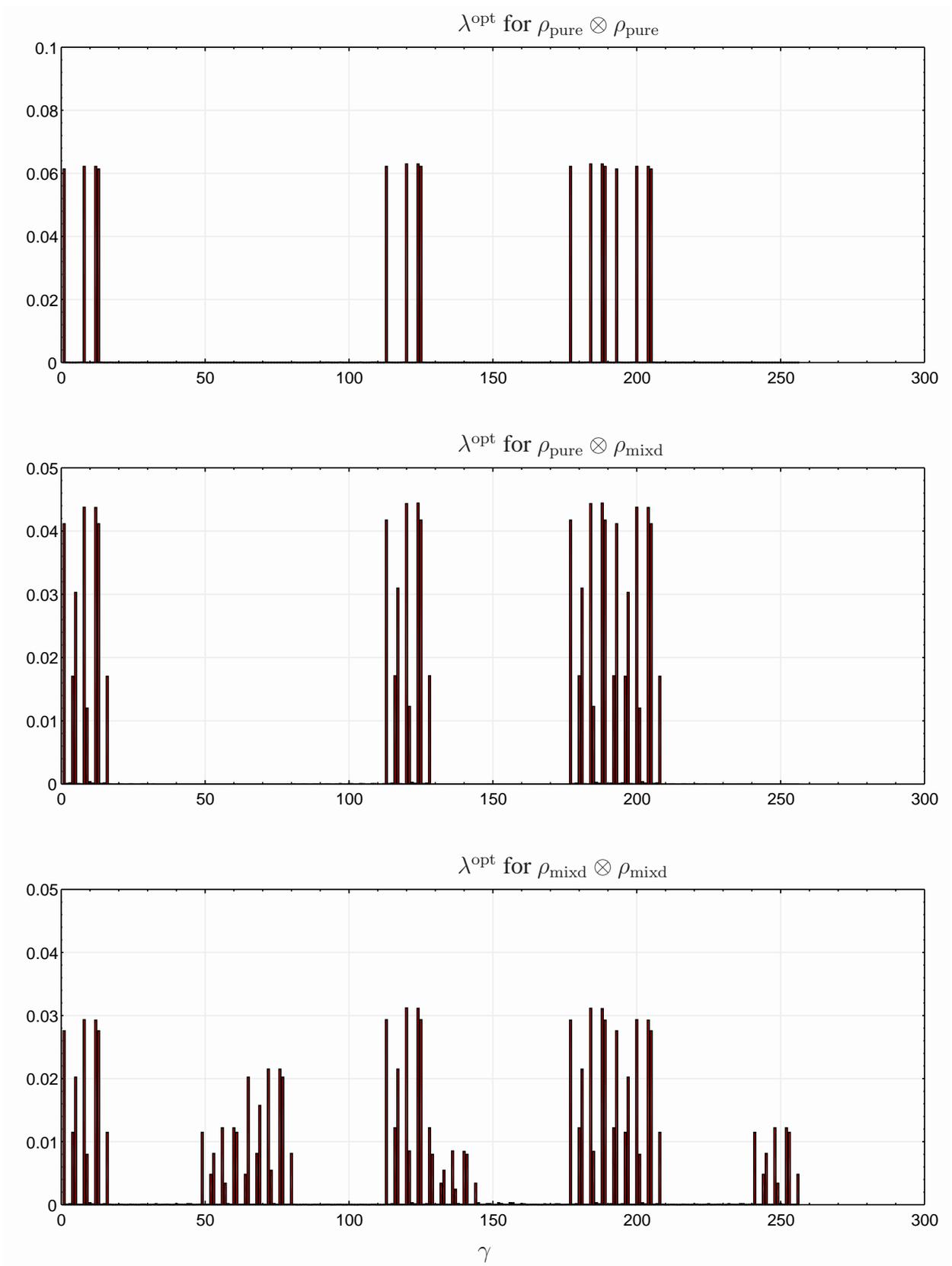}
\caption{optimal distributions -- two-arms}
\label{fig:lamopt2arm}
\end{figure}

Figure \reffig{lamopt2arm} shows the optimal distributions $\lamh$
versus configurations $\gam=1,\ldots,256$ for all three input states.
In this case the optimal distributions $\lamh$ are nearly uniform in
magnitude.  The following table compares the distributions for
optimal, suboptimal with 25 angles, uniform at the 25 suboptimal
angles, and uniform at all 256 angles by examining the minimum number
of experiments required to obtain an RMS estimation error of no more
than 0.01.
\[
\renewcommand{\arraystretch}{1}
\bea{|c||c|c|c|c|}
\hline
\mbox{input state} 
& 
\bea{c}
\mbox{optimal}
\\
\round{\nex \lamopt}
\\
\nconf=256
\eea
& 
\bea{c}
\mbox{suboptimal}
\\
\round{\nex\lamsub}
\\
\nconf=25
\eea
&
\bea{c}
\mbox{uniform}
\\ 
(1/25)\ {\bf 1}(\lamsub)
\\
\nconf=25
\eea
&
\bea{c}
\mbox{uniform}
\\ 
(1/256)\ {\bf 1}(\lamopt)
\\
\nconf=256
\eea
\\
\hline
\rhopur\otimes\rhopur & 81,870 & 81,877 & 86,942 & 115,165\\
\rhopur\otimes\rhomix & 135,158 & 139,151 & 156,219 & 226,234\\
\rhomix\otimes\rhomix & 225,739 & 288,163 & 262,833 & 427,292\\
\hline
\eea
\]
As might be expected it is easier to estimate all pure states than
mixed states. The table also shows that the optimal solution can be
used effectively to guide the selection of suboptimal distributions.

\psfrag{gam}{$\gam$}
\psfrag{r0101}{$\lamopt$ for $\rhopur\otimes\rhopur$}
\psfrag{r0102}{$\lamopt$ for $\rhopur\otimes\rhomix$}
\psfrag{r0202}{$\lamopt$ for $\rhomix\otimes\rhomix$}

\clearpage
\subsection{Maximum likelihood state distribution estimation}
\label{sec:dist est}

\begin{figure}[h]
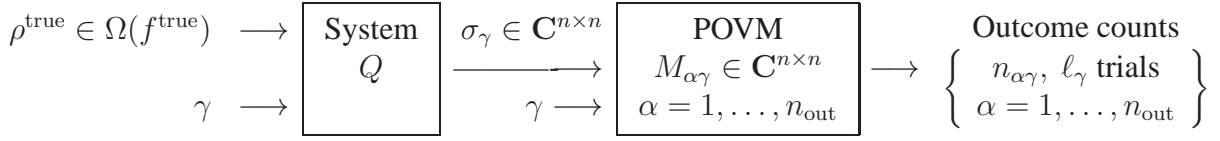

\[
\bea{rr}
\rhotrue\in\Om(\ftrue) & \longrightarrow
\\
\mbox{} & \mbox{}
\\
\gam & \longrightarrow
\eea
\mathbox{
\bea{c}
\mbox{System}
\\
\qsys
\\
\mbox{}
\eea
}
\bea{r}
\sig_\gam\in\Cbfnn 
\\ 
\mbox{----------}\!\!\!\longrightarrow 
\\ 
\gam \longrightarrow
\eea
\mathbox{
\bea{c}
\mbox{POVM}
\\
M_{\alf\gam} \in\Cbfnn
\\
\alf=1,\ldots,\nout
\eea
}
\longrightarrow
\bea{c}
\mbox{Outcome counts}
\\
\seq{
\bea{c} n_{\alf\gam},\ \ell_\gam~\mbox{trials} 
\\ \alf=1,\ldots,\nout \eea
}
\eea
\]
\caption{System/POVM.}
\label{fig:dist}
\end{figure}

A variation on the state estimation problem is to estimate the {\em
distribution} of a known set of input states.  The set-up for data
collection is shown schematically in Figure \ref{fig:dist} for
configuration $\gam$.

\noindent
In this case the input state $\rhotrue$ is drawn from,
\beq[eq:omin]
\Om(\ftrue)
=
\set{\rho(i),\ \ftrue(i)}
{i=1,\ldots,\nin}
\eeq
which consists of a set of known states,
$\seq{\rho(1),\ldots,\rho(\nin)}$, with corresponding unknown
occurrence probabilities $\ftrue=\seq{\ftrue(1),\ldots,\ftrue(\nin)}$
where $0\leq\ftrue(i)\leq 1,\forall i$ and $\sum_i \ftrue(i)=1$. The
objective is to use the data and knowledge of the input state set to
estimate the vector of occurrence probabilities $\ftrue$. Proceeding
as before, assume that the input state {\em model} is $\rho\in\Om(f)$
where $\Om(f)=\set{\rho(i),\ f(i)} {i=1,\ldots,\nin} $ and where the
vector of distributions $f=[f(1),\ldots,f(\nin)]^T\in\Rbf^\nin$ is to
be estimated.  In this case the input state can be represented by the
mixed state,
\beq[eq:rhof]
\rho(f) = \sum_{i=1}^\nin\ f(i) \rho(i)
\eeq
The ML estimate of $f$ is then the solution of the optimization
problem:
\beq[eq:maxlike f]
\bea{ll}
\mbox{minimize} 
& 
L(f)=
-\sum_{\alf,\gam}
n_{\alf\gam}
\log\trace\ O_{\alf\gam}\rho(f)
\\
\mbox{subject to} 
& 
\sum_{i=1}^\nin f(i)=1,
\;\;
f(i)\geq 0,\ i=1,\ldots,\nin
\eea
\eeq
As in the MLE for quantum state estimation \refeq{maxlike1}, the
objective is log-convex in the state $\rho(f)$, the state is linear in
$f\in\Rbf^\nin$, and the constraints form a convex set in $f$. Hence,
this is a convex optimization problem in the variable $f$.  Combining
\refeq{rhof} with \refeq{maxlike f} gives the more explicit form,
\beq[eq: maxlike f]
\bea{ll}
\mbox{minimize} 
& 
L(f)=
-\sum_{\alf,\gam}
n_{\alf\gam}
\log a_{\alf\gam}^T f
\\
\mbox{subject to} 
& 
\sum_{i=1}^\nin f(i)=1,
\;\;
f(i)\geq 0,\ i=1,\ldots,\nin
\\
&
a_{\alf\gam}
=
\left[
\trace\ O_{\alf\gam}\rho(1)
\cdots
\trace\ O_{\alf\gam}\rho(\nin)
\right]^T\in\Rbf^\nin,\;
\forall \alf,\gam
\eea
\eeq
Here again as in \refeq{lsopt cvx} we could solve for $f$ using the
empirical estimate of the outcome probabilities as in \refeq{lsopt}:
\beq[eq:lsopt f]
\bea{ll}
\mbox{minimize}
&
\sum_{\alf,\gam}\ 
w_\gam
\left( \pemp_{\alf\gam} - a_{\alf\gam}^T f \right)^2
\\
\mbox{subject to}
&
\sum_{i=1}^\nin f(i)=1,
\;\;
f(i)\geq 0,\ i=1,\ldots,\nin
\eea
\eeq
%

\subsection{Experiment design for state distribution estimation}
\label{sec:expdes qs dist}

Let $\fsurr\in\Rbf^\nin$ be a surrogate for the true state
distribution, $f^\true$. Following the derivation of \refeq{var rho}
in Appendix \refsec{var rho}, the associated (relaxed) optimal
experiment design problem is,
\beq[eq:expdes qs dist]
\bea{ll}
\mbox{minimize}
&
V(\lam,\fsurr)=
\trace\left( 
\sumgam\ \lam_\gam G_\gam(\fsurr)
\right)^{-1}
\\
\mbox{subject to}
&
\sumgam\ \lam_\gam = 1
\\
&
\lam_\gam\geq 0,\;
\gam=1,\ldots,\nconf
\eea
\eeq
where 
\beq[eq:expdes qs 1] 
\bea{rcl}
G_\gam(\fsurr)
&=&
\ds
\Ceq^T
\left(
\sumalf\
\frac{a_{\alf\gam}a_{\alf\gam}^T}
{p_{\alf\gam}(\fsurr)}
\right)
\Ceq
\in\Rbf^{\nin-1 \times \nin-1}
\eea
\eeq
with $a_{\alf\gam}\in\Rbf^\nin$ from \refeq{maxlike f} and where
$\Ceq\in\Rbf^{\nin\times\nin-1}$ is part of the unitary matrix $W$ in
the singular value decomposition: $1_{\nin}^T=USW^T,\ W=[c\
\Ceq]\in\Rbf^{\nin\times\nin}$.

\section{Quantum Process Tomography: OSR Estimation}
\label{sec:qptosr}

\begin{figure}[h]
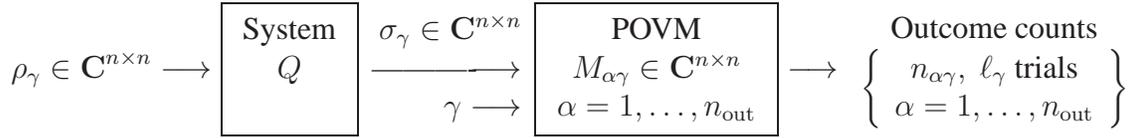

\[
\rho_\gam\in\Cbfnn \longrightarrow
\mathbox{
\bea{c}
\mbox{System}
\\
\qsys
\\
\mbox{}
\eea
}
\bea{r}
\sig_\gam\in\Cbfnn 
\\ 
\mbox{----------}\!\!\!\longrightarrow 
\\ 
\gam \longrightarrow
\eea
\mathbox{
\bea{c}
\mbox{POVM}
\\
M_{\alf\gam} \in\Cbfnn
\\
\alf=1,\ldots,\nout
\eea
}
\longrightarrow
\bea{c}
\mbox{Outcome counts}
\\
\seq{
\bea{c} n_{\alf\gam},\ \ell_\gam~\mbox{trials} 
\\ \alf=1,\ldots,\nout \eea
}
\eea
\]
\caption{System/POVM.}
\label{fig:qpt}
\end{figure}

We explore two methods of identification for determining the
$\qsys$-system from data: (i) in this section, estimating the OSR in a
fixed basis, and (ii) in Section \refsec{ham est}, estimating
Hamiltonian parameters.  In either case, the set-up is now as shown in
Figure \ref{fig:qpt}.

\noindent
The main difference between state tomography (Figure \ref{fig:rho})
and process tomography (Figure \ref{fig:qpt}) is that in the latter
case the input state is prepared at specific values, $\rho_\gam$,
depending on the configuration, whereas the $\qsys$-system does not
depend on the configuration. If the process varies with every change
in configuration it would be much more difficult to estimate; a model
of configuration dependence would need to be established. This
situation is perhaps amenable with Hamiltonian parameter estimation
but will not be pursued here.

\subsection{Maximum likelihood OSR estimation}
\label{sec:qpt osr}

As already stated, the Krause matrices for modeling the (trace
preserving) $\qsys$-system in this case are {\em not} dependent on the
configuration $\gam$, specifically, $K=\set{K_k}{k=1,\ldots,\kappa}$
with $\kappa \leq n^2$.  Using \refeq{osr}, the reduced state model in
Figure \ref{fig:qpt} as a function of $K$ is,
\beq[eq:osr est]
\sig_\gam(K)
=\qsys(\rho_\gam,K)
=
\sum_{k=1}^\kappa\ K_k \rho_\gam K_k^*,
\;\;\;
\sum_{k=1}^\kappa\ K_k^* K_k = I_n
\eeq
Combining the above with the measurement model \refeq{povm} gives the
probability outcomes model,
\beq[eq:pout osr est]
p_{\alf\gam}(K)
=
\trace\ O_{\alf\gam}(K)\rho_\gam,
\;\;
O_{\alf\gam}(K) 
=
\sum_{k=1}^\kappa\ K_k^* M_{\alf\gam} K_k
\eeq
The log-likelihood function \refeq{loglike1} is,
\beq[eq:loglike osr]
L(D,K) 
= 
-\sum_{\alf,\gam}\ 
n_{\alf\gam} \log \trace\ O_{\alf\gam}(K)\rho_\gam
\eeq  
An ML estimate of $K$ is then a solution to,
\beq[eq:mlosr k]
\bea{ll}
\mbox{minimize}
&
L(D,K)
=
-\sum_{\alf,\gam}\ 
n_{\alf\gam} \log \trace\ 
\sum_{k=1}^\kappa\ K_k^* M_{\alf\gam} K_k
\rho_\gam
\\
\mbox{subject to}
&
\sum_{k=1}^\kappa\ K_k^* K_k = I_n
\eea
\eeq
This is not a convex optimization for two reasons: the equality
constraint is not linear in $K$ and the objective function is not
convex.  The problem can be transformed -- more accurately, embedded
-- into a convex optimization problem by expanding the Kraus matrices
in a fixed basis. The procedure, described in \cite[\S
8.4.2]{NielsenC:00}, is as follows: since any matrix in $\Cbfnn$ can
be represented by $n^2$ complex numbers, let
\beq[eq:basis]
\set{B_i\in\Cbfnn}{i=1,\ldots,n^2}
\eeq
be a basis for matrices in $\Cbfnn$. The Kraus matrices can thus be
expressed as,
\beq[eq:krausb]
K_k = \sum_{i=1}^{n^2}\ a_{ki} B_i,\;
k=1,\ldots,\kappa
\eeq
where the $n^2$ coefficients $\seq{a_{ki}}$ are complex scalars.
Introduce the matrix $X\in\Cbf^{n^2\times n^2}$, often referred to as
the {\em superoperator}, with elements,
\beq[eq:osr2]
X_{ij} = \sum_{k=1}^\kappa\ a_{ki}^* a_{kj},
\;\;\;
i,j=1,\ldots,n^2
\eeq
As shown in \cite{NielsenC:00}, from the requirement to preserve
probability, $X$ is restricted to the convex set,
\beq[eq:xset]
X \geq 0,
\;\;\;
\sum_{i,j=1}^{n^2}\ X_{ij}\ B_i^* B_j = I_n
\eeq
The system output state \refeq{osr est} and outcome probabilities
\refeq{pout osr est} now become,
\beq[eq:osr1]
\bea{rcl}
\sig_\gam(X)
&=&
\ds
\qsys(\rho_\gam,X)
=
\sum_{i,j=1}^{n^2}\ X_{ij}\ B_i \rho_\gam B_j^*
\\&&\\
p_{\alf\gam}(X)
&=&
\ds
\trace\ O_{\alf\gam}\ Q(\rho_\gam,X)
=
\trace\ X R_{\alf\gam}
\eea
\eeq
where the matrix $R_{\alf\gam}\in\Cbf^{n^2\times n^2}$ has elements,
\beq[eq:rmat]
[R_{\alf\gam}]_{ij}
=
\trace\ B_j \rho_\gam B_i^* O_{\alf\gam},\;
i,j = 1,\ldots,n^2
\eeq
Quantum process tomography is then estimating $X\in\Cbf^{n^2\times
n^2}$ from the data set $D$ \refeq{data}.  An ML estimate is obtained
by solving for $X$ from:
\beq[eq:mle x]
\bea{ll}
\mbox{minimize}
&
L(D,X) 
= 
-\sum_{\alf,\gam}\
n_{\alf\gam} \log \trace\ X R_{\alf\gam}
\\
\mbox{subject to}
&
X \geq 0,
\;\;\;\;
\sum_{ij}\ X_{ij}\ B_i^*B_j = I_n
\eea
\eeq
This problem has essentially the same form as \refeq{maxlike1}, and
hence is also a convex optimization problem with the optimization
variables being the elements of the matrix $X$. Since
$X=X^*\in\Cbfntnt$, it can be parametrized by $n^4$ real
variables. Accounting for the $n^2$ real linear equality constraints,
the number of free (real) variables in $X$ is thus $n^4-n^2$. This can
be quite large even for a relatively small number of qubits, \eg, for
$q=[1,\ 2,\ 3,\ 4]$ qubits, $n=2^q=[2,\ 4,\ 8,\ 16]$ and
$n^4-n^2=[12,\ 240,\ 4032,\ 65280]$. This exponential (in qubit)
growth is the main drawback to using this approach.

The $X$ (superoperator) matrix can be transformed back to Kraus
operators via the singular value decomposition \cite[\S
8.4.2]{NielsenC:00}.  Specifically, let $X=VSV^*$ with unitary $V\in
\Cbf^{n^2\times n^2}$ and $S={\rm diag}(s_1\ \cdots\ s_{n^2})$ with
the singular values ordered so that $s_1 \geq s_2 \geq\ \cdots\ \geq
s_{n^2} \geq 0$. Then the coefficients in the basis representation of
the Kraus matrices \refeq{krausb} are,
\beq[eq:aki]
a_{ki} = \sqrt{s_k}\ V_{i k}^*,\
k,i=1,\ldots,n^2
\eeq
Theoretically there can be fewer then $n^2$ Kraus operators. For
example, if the $\qsys$ system is unitary, then,
\beq[eq:u rho u]
\qsys(\rho) = U \rho U^*
\eeq
In effect, there is one Kraus operator, $U$, which is unitary and of
the same dimension as the input state $\rho$.  The corresponding $X$
matrix is a dyad, hence $\rank\ X=1$. A rank constraint is not
convex. However, the $X$ matrix is symmetric and positive
semidefinite, so the heuristic from \cite{FazelHB:01} applies where the
rank constraint is replaced by the trace constraint,
\beq[eq:rankx]
\trace\ X \leq \eta
\eeq
From the singular value decomposition of $X$, $\trace\ X = \sum_k
s_k$, and hence, adding the constraint \refeq{rankx} to \refeq{mle x}
will force some (or many) of the $s_k$ to be small which can be
eliminated (post-optimization) thereby reducing the rank. The
auxiliary parameter $\eta$ can be used to find a tradeoff between
simpler realizations and performance. The estimation problem is then:
\beq[eq:mle x eta]
\bea{ll}
\mbox{minimize}
&
L(D,X) 
= 
-\sum_{\alf,\gam}\
n_{\alf\gam} \log \trace\ X R_{\alf\gam}
\\
\mbox{subject to}
&
X \geq 0,
\;\;\;\;
\sum_{ij}\ X_{ij}\ B_i^*B_j = I_n
\\
&
\trace\ X \leq \eta
\eea
\eeq
%

\subsection{Experiment design for OSR estimation}
\label{sec:expdes osr}

Let $\Xsurr\in\Cbfntnt$ be a surrogate for the true OSR, $X^\true$. As
derived in Appendix \refsec{expdes osr der}, the associated (relaxed)
optimal experiment design problem is,
\beq[eq:expdes osr]
\bea{ll}
\mbox{minimize}
&
V(\lam,\Xsurr)=
\trace\left( 
\sumgam\ \lam_\gam G_\gam(\Xsurr)
\right)^{-1}
\\
\mbox{subject to}
&
\sumgam\ \lam_\gam = 1
\\
&
\lam_\gam\geq 0,\;
\gam=1,\ldots,\nconf
\eea
\eeq
where 
\beq[eq:expdes osr 1] 
\bea{rcl}
G_\gam(\Xsurr)
&=&
\ds
\Ceq^*
\left(
\sumalf\
\frac{a_{\alf\gam} a_{\alf\gam}^* }{p_{\alf\gam}(\Xsurr)} 
\right)
\Ceq
\\&&\\
a_{\alf\gam} 
&=&
\ds
\vec\ R_{\alf\gam} \in \Cbf^{n^4} 
\eea
\eeq
and $\Ceq\in\Cbf^{n^4\times n^4-n^2}$ is part of the unitary matrix
$W=[C\ \Ceq]\in\Cbf^{n^4\times n^4}$ in the singular value
decomposition of the $n^2\times n^4$ matrix,
\beq[eq:expdes osr 2]
\left[a_1\ \cdots\ a_{n^4}\right]
=
U \left[ \sqrt{n}I_{n^2}\;\; 0_{n^2\times n^4-n^2} \right] W^*
\eeq
with with $a_k=\vec(B_i^*B_j)\in\Cbf^{n^2}$ for $k=i+(j-1)n^2,\
i,j=1,\ldots,n^2$.  The columns of $\Ceq$, \ie, the last $n^4-n^2$
columns of $W$, are a basis for the nullspace of $\left[a_1\ \cdots\
a_{n^4}\right]$.

\subsection{Example: experiment design for OSR estimation}
\label{sec:ex expdes osr}

Consider the POVM set from the one-arm photon detector
(\refsec{example qs}) using all combinations of the following set
wave-plate angles,
\[
h=[0\ 30\ 45],\; q=[0\ 30\ 45]
\]
Assume detector efficiency $\eta=0.75$ and dark count probability
$\del=0.05$. The set of inputs (state configurations) is 
\[
\ketlo,\; \kethi,\; \ketp=(\ketlo+\kethi)/\sqrt{2},\; 
\ketm=(\ketlo+i\kethi)/\sqrt{2}
\]
The 9 combinations of angles together with the 4 combinations of input
states gives a total of 36 configurations, $\gam=1,\ldots,\nconf=36$.

\begin{figure}[h]
\centering
\epsfysize=3.5in
\epsfxsize=6.5in
\epsfboxit{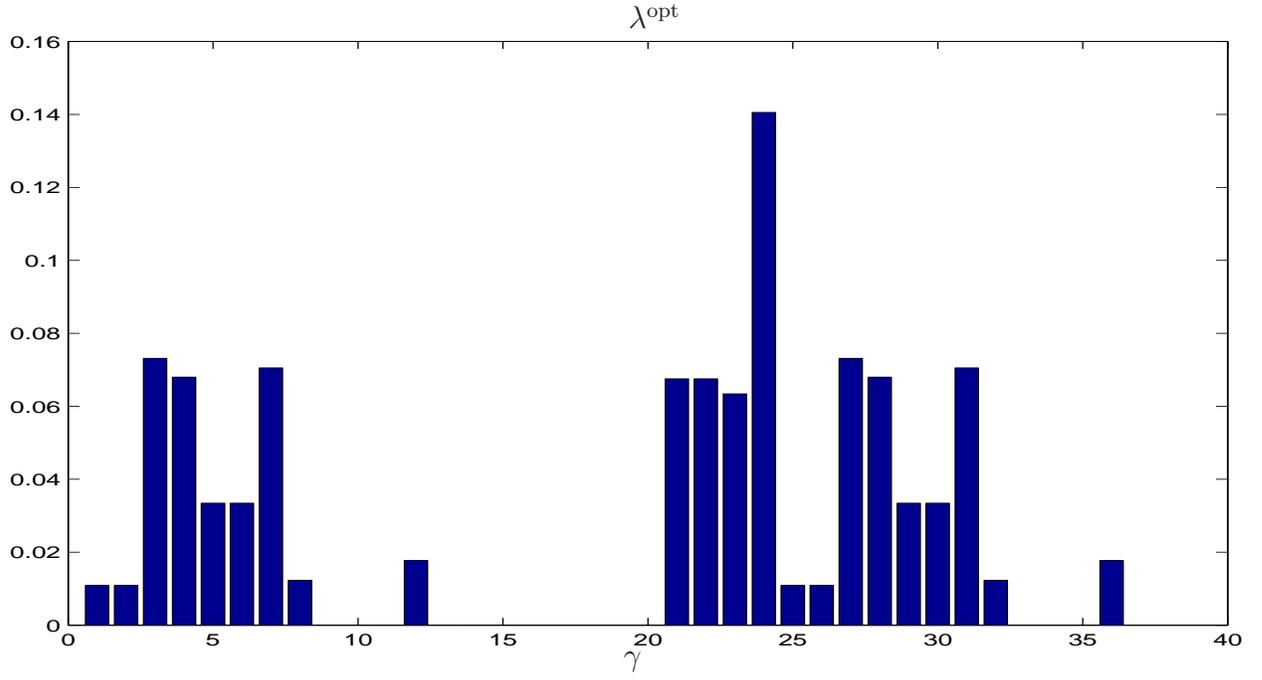}
\caption{optimal distributions for OSR estimation}
\label{fig:lamopt_qp}
\end{figure}

Figure \ref{fig:lamopt_qp} shows the optimal distribution of
experiments for the 36 configurations using the true OSR corresponding
to the Pauli basis set 
$
\seq{
I_2/\sqrt{2},\ 
\sig_x/\sqrt{2},\ 
\sig_y/\sqrt{2},\ 
\sig_z/\sqrt{2}
}
$.
Since the system is simply the identity, $\qsys(\rho)=\rho$, with this
basis choice, $X^\true=\diag(2\ 0\ 0\ 0)$.  (No knowledge of the
system being identity is used, hence, all elements of $X^\true$ are
estimated, not just the single element in the ``11'' location.)

The following table displays the minimum number of experiments
required to meet estimation accuracies of 0.05 and 0.01 for both
uniform and optimal distributions.
\[
\bea{|c||c|c|}
\hline
\mbox{accuracy} & \lam^\opt & \lam^{\rm unif}
\\
\hline
0.01 & 856,676 & 1,304,561\\
0.05 & 34,268 & 52,183\\
\hline
\eea
\]
Approximately 35\% fewer experiments are needed using the optimal
distribution. Although not dramatic, as in the photon estimation
example \refsec{example qs}, there is a large penalty, in terms of
time, for changing wave-plate angles.

\psfrag{gam}{$\gam$}
\psfrag{lam}{$\lamopt$}

The following Table shows the 22 out of 36 configurations for
$\lam^\opt_\gam > 0.01$.
\[
\bea{|c|c|c|c|c|}
\hline
\gam & \lam^\opt_\gam & h_\gam & q_\gam & \rho_\gam
\\
\hline
1 & 0.011 & 0 & 0 & \ketlo \\   
2 & 0.011 & 0 & 0 & \kethi \\   
3 & 0.073 & 0 & 0 & \ketp \\   
4 & 0.068 & 0 & 0 & \ketm \\   
5 & 0.033 & 0 & 30 & \ketlo \\  
6 & 0.033 & 0 & 30 & \kethi \\  
7 & 0.071 & 0 & 30 & \ketp \\  
8 & 0.012 & 0 & 30 & \ketm \\  
12 & 0.018 & 0 & 45 & \ketm \\ 
21 & 0.068 & 30 & 45 & \ketlo \\
22 & 0.068 & 30 & 45 & \kethi \\
23 & 0.063 & 30 & 45 & \ketp \\
24 & 0.141 & 30 & 45 & \ketm \\
25 & 0.011 & 45 & 0 & \ketlo \\ 
26 & 0.011 & 45 & 0 & \kethi \\ 
27 & 0.073 & 45 & 0 & \ketp \\ 
28 & 0.068 & 45 & 0 & \ketm \\ 
29 & 0.033 & 45 & 30 & \ketlo \\
30 & 0.033 & 45 & 30 & \kethi \\
31 & 0.071 & 45 & 30 & \ketp \\
32 & 0.012 & 45 & 30 & \ketm \\
36 & 0.018 & 45 & 45 & \ketm \\
\hline
\eea
\]

\subsection{Maximum likelihood OSR distribution estimation}
\label{sec:osr dist}

Suppose the Kraus matrices are known up to a scale factor which is
related to its probability of occurrence, that is,
\beq[eq:osr dist]
\bea{lll}
K_k = \sqrt{q_k}\ \Kb_k
&
\sum_{k=1}^\kappa q_k \Kb_k^* \Kb_k=I_n
&
{}
\\
\sum_{k=1}^\kappa q_k=1
&
q_k \geq 0
&
k=1,\ldots,\kappa
\eea
\eeq
One interpretation of this system model is that one of the matrices,
say $\Kb_1$, is the nominal (unperturbed) system, and the others,
$\Kb_k,k=2,\ldots,\kappa$, are perturbations, each of them occurring
with probability $q_k$. Examples of perturbations include the typical
errors which can be handled by quantum error correction codes, \eg,
depolarization, phase damping, phase and bit flip; see, \eg,
\cite[Ch.8]{NielsenC:00}.

The goal is to use the data to estimate the unknown vector of
probabilities, $q=[q_1 \cdots q_\kappa]^T \in\Rbf^\kappa$. Using the
system model \refeq{osr dist}, the model probability outcomes are,
\beq[eq:pout osr dist]
\bea{rcl}
p_{\alf\gam} 
&=&
\trace\ M_{\alf\gam} \sum_{k=1}^\kappa
q_k\ \Kb_k \rho_\gam \Kb_k^*
=
a_{\alf\gam}^T q
\\
a_{\alf\gam} 
&=&
\left[
\trace\ M_{\alf\gam} \Kb_1 \rho_\gam \Kb_1^*
\cdots
\trace\ M_{\alf\gam} \Kb_\kappa \rho_\gam \Kb_\kappa^*
\right]^T
\in\Rbf^\kappa
\eea
\eeq
The ML estimate of $q\in\Rbf^\kappa$ is the solution of the
optimization problem,
\beq[eq:mle osr dist]
\bea{ll}
\mbox{minimize} 
& 
L(q)=
-\sum_{\alf,\gam}
n_{\alf\gam}
\log a_{\alf\gam}^T q
\\
\mbox{subject to} 
& 
\sum_{k=1}^\kappa q_k=1,
\;\;
q_k\geq 0,\ k=1,\ldots,\kappa
\eea
\eeq
This is a convex optimization problem and is essentially in the same
form as problem \refeq{maxlike f} which seeks the ML estimate of the
input state distribution.

\subsection{Experiment design for OSR distribution estimation}
\label{sec:expdes osr dist}

The formulation here is directly analogous to that of experiment
design for state distribution estimation \refsec{expdes qs dist}.
Let $\qsurr\in\Rbf^\nin$ be a surrogate for the true OSR distribution,
$q^\true$. Following the lines of the derivation in Appendix
\refsec{expdes osr der}, the associated (relaxed) optimal
experiment design problem is,
\beq[eq:expdes osr dist]
\bea{ll}
\mbox{minimize}
&
V(\lam,\qsurr)=
\trace\left( 
\sumgam\ \lam_\gam G_\gam(\qsurr)
\right)^{-1}
\\
\mbox{subject to}
&
\sumgam\ \lam_\gam = 1
\\
&
\lam_\gam\geq 0,\;
\gam=1,\ldots,\nconf
\eea
\eeq
where 
\beq[eq:expdes qs 2] 
\bea{rcl}
G_\gam(\qsurr)
&=&
\ds
\Ceq^T
\left(
\sumalf\
\frac{a_{\alf\gam}a_{\alf\gam}^T}
{p_{\alf\gam}(\qsurr)}
\right)
\Ceq
\in\Rbf^{\nin-1 \times \nin-1}
\eea
\eeq
with $a_{\alf\gam}\in\Rbf^\nin$ from \refeq{mle osr dist} and where
$\Ceq\in\Rbf^{\nin\times\nin-1}$ is part of the unitary matrix $W$ in
the singular value decomposition: $1_{\nin}^T=USW^T,\ W=[c\
\Ceq]\in\Rbf^{\nin\times\nin}$.

\subsection{Example: experiment design for OSR distribution estimation}
\label{sec:ex osr dist}

Consider a quantum process, or channel, where a single qubit state,
$\rho\in\Cbf^{2\times 2}$, is corrupted by a {\em bit-flip error} with
occurrence probability $q_B$ and a {\em depolarizing error} with
occurrence probability $q_D$. The process is described by the quantum
operation,\footnote{
$X$ is one of the three $2\times 2$ Pauli spin matrices:
$
X = \mattwo{0}{1}{1}{0},\
Y = \mattwo{0}{-i}{i}{0},\
Z = \mattwo{1}{0}{0}{-1}
$
}
\beq[eq:bf dep]
\bea{l}
Q(\rho,q) 
=
\ds
q_I\ \rho + q_B\ X\rho X
+
q_D\ I/2
\\
\\
q_I+q_B+q_D = 1
\eea
\eeq
where $q_I = 1 -(q_B+q_D)$ is the probability of no error occurring.
The probability of observing outcome $\alf$ with the system in
configuration $\gam$ is,\footnote{
As shown in \cite[\S 8.3]{NielsenC:00}, an equivalent set of OSR
elements which describe \refeq{bf dep} are,
$
\sqrt{q_I}\ I,\
\sqrt{q_B}\ X,\
\sqrt{q_D}\ I/2,\ 
\sqrt{q_D}\ X/2,\ 
\sqrt{q_D}\ Y/2,\ 
\sqrt{q_D}\ Z/2
$.
Forming the probability outcomes in terms of this expansion
results in an overparamtrization.
}
\beq[eq:palfgam q]
\bea{rcl}
p_{\alf\gam}(q)
&=&
\trace\ M_{\alf\gam} Q(\rho_\gam,q)
\\
&=&
\left[
\bea{ccc}
\trace\ M_{\alf\gam}\rho_\gam
&
\trace\ M_{\alf\gam} X\rho_\gam X
&
\trace\ M_{\alf\gam}/2
\eea
\right]
\
\left[
\bea{c} q_I \\ q_B \\ q_D \eea
\right]
\\
&=&
a_{\alf\gam}^T\ q
\eea
\eeq
An interesting aspect of this problem is that not all input states
$\rho_\gam$ lead to identifiability of the occurrence probabilities. And
this is independent of the choice of POVM $M_{\alf\gam}$. To see this
consider the single pure input state,
\beq
\rho_\gam=\psi\psi^*,
\;
\psi=\left[\bea{c} a \\ b \eea\right],
\;
|a|^2 + |b|^2 =1
\eeq
The output of the channel \refeq{bf dep} is then,
\beq
Q(\psi\psi^*,q)
=
\left[
\bea{cc}
q_I |a|^2 + q_B |b|^2 + q_D/2
&
q_I ab^* + q_B a^*b
\\
q_I a^*b + q_B ab^*
&
q_I |b|^2 + q_B |a|^2 + q_D/2
\eea
\right]
\eeq
Suppose we knew the elements of $Q(\psi\psi^*,q)$ perfectly; call them
$Q_{11},Q_{12},Q_{22}$. Then in principal we could solve for the three
occurrence probabilities from the linear system of equations,
\beq
\underbrace{
\left[
\bea{ccc}
|a|^2 & |b|^2 & 1/2
\\
|b|^2 & |a|^2 & 1/2
\\
ab^* & a^*b & 0
\eea
\right]
}_{R}
\
\left[
\bea{c} q_I \\ q_B \\ q_D \eea
\right]
=
\left[
\bea{c} Q_{11} \\ Q_{22} \\ Q_{12} \eea
\right]
\eeq
If $\det\ R=0$ then no unique solution exists; the occurrence
probabilities are not {\em identifiable}. Specifically, $\det\ R=0$
for all $a,b\in\Cbf$ such that,
\beq
(\real\ ab^*)(|b|^2-|a|^2) = 0,
\;
|a|^2+|b|^2=1
\eeq
Equivalently, $\det\ R=0$ for the following sets of $a,b\in\Cbf$:
\beq[eq:bad ab]
(a=0,\ |b|=1),
\;
(|a|=1,\ b=0),
\;
(|a| = |b| = 1/\sqrt{2})
\eeq
Let the input state be a single pure state of the form
\beq
\psi(\th)=\left[\bea{c} \cos\th \\ \sin\th \eea\right],
\eeq
Suppose the angle $\th$ is restricted to the range $0 \leq \th \leq
90^\circ$. Using \refeq{bad ab}, the occurrence probabilities are not
identifiable for the angles $\th$.  and respectively, the states
$\psi(\th)$, in the sets,
\beq[eq:bad th]
\th\in\seq{0,\ 45^\circ,\ 90^\circ},
\;
\psi(\th)\in
\seq{
\left[\bea{c} 1 \\ 0 \eea\right],\
\left[\bea{c} 1/\sqrt{2} \\ 1/\sqrt{2} \eea\right],\
\left[\bea{c} 0 \\ 1 \eea\right]
}
\eeq
Unfortunately, this excludes inputs identical to the computational
basis states $\ket{0}$ or $\ket{1}$, respectively, $\psi(\th)$ with
$\th=0$ or $\th=90^\circ$. 

We now solve the (relaxed) experiment design problem \refeq{expdes osr
dist} for occurrence probabilities $(q_I,\ q_B,\ q_D)=(0.6,\ 0.2,\
0.2)$ and with the POVM set given by \refeq{povm hqgam}.  For
illustrative purposes, we use only 16 of the 100 configurations
represented by the wave plate angles \refeq{hq1arm}. Specifically, the
wave-plate angles are:
$
\seq{h=0,\ 15,\ 30,\ 45} \times
\seq{q=0,\ 15,\ 30,\ 45}
$.
The optimization results are presented in the following table which
shows the number of experiments per configuration (each of the 16
angle pairs) required to achieve an accuracy of 0.01 for the input
states corresponding to the angles $\th\in\seq{2,\ 10,\ 25,\ 35,\
44}$.

\bc
\btab{|c|c||r|r|r|r|r|}
\hline
\mcol{2}{|c||}{configurations} & \mcol{5}{|c|}{experiments per configuration}\\
\hline\  
$h$ & $q$ & $\th=2$ & $\th=10$ & $\th=25$ & $\th=35$ & $\th=44$ \\
\hline\hline

0 & 0 & 62,244 & 15,453 & 13,386 & 31,275 & 2,136,560 \\  
0 & 15 & 1 & 1 & 1 & 1 & 1 \\                       
0 & 30 & 1 & 0 & 0 & 0 & 1 \\                       
0 & 45 & 1 & 0 & 0 & 0 & 1 \\                       
15 & 0 & 1 & 0 & 0 & 0 & 1 \\                       
15 & 15 & 1 & 1 & 1 & 1 & 1 \\                      
15 & 30 & 73,096 & 11,277 & 4,765 & 9,006 & 107,371 \\   
15 & 45 & 2,080,984 & 89,588 & 18,598 & 14,573 & 62,187 \\
30 & 0 & 1 & 0 & 0 & 0 & 1 \\                       
30 & 15 & 1 & 0 & 0 & 0 & 1 \\                      
30 & 30 & 1 & 0 & 0 & 0 & 1 \\                      
30 & 45 & 2,080,984 & 89,588 & 18,598 & 14,573 & 62,187 \\
45 & 0 & 62,244 & 15,453 & 13,386 & 31,275 & 2,136,560 \\ 
45 & 15 & 1 & 1 & 1 & 1 & 1 \\                      
45 & 30 & 1 & 0 & 0 & 0 & 1 \\                      
45 & 45 & 1 & 0 & 0 & 0 & 1 \\      
\hline\hline                           
\mcol{2}{|c||}{$\nex\ =$} & 4,359,563 & 221,362 & 68,736 & 100,705 & 4,504,876 \\
\hline
\etab
\ec

The numerical example shows that for input states close to those
states which make the problem not identifiable \refeq{bad th}, the
number of experiments required to achieve the specified accuracy grows
very large.  In this case, $\th=2$ and $\th=44$ are close to the bad
angles 0 and 45, and the number of experiments is quite large.

\section{Hamiltonian Parameter Estimation}
\label{sec:ham est}

The process of modeling a quantum system in this case begins with the
construction of a Hamiltonian {\em operator} on an infinite
dimensional Hilbert space. {\em Eventually}, a finite dimensional
approximation is invoked in order to calculate anything. (In some
cases a finite dimensional model is immediately appropriate, \eg, spin
systems, \cite[Ch.12-9]{FeynmanLS:65}.) The finite dimensional model
is the starting point here.

\subsection{Maximum likelihood Hamiltonian parameter estimation}
\label{sec:mle hampar}

The quantum system is modeled by a finite dimensional Hamiltonian
matrix $H(t,\th)\in\Cbfnn$, having a known dependence on time $t,\
0\leq t\leq t_f$, and on an unknown parameter vector
$\th\in\Rbf^{\nth}$. The model density matrix will depend on $\th$ and
the initial (prepared and known) state drawn from the set of states
$\set{\rho^\init_\bet\in\Cbfnn}{\bet=1,\ldots,\nin}$.  Thus, the
density matrix associated with initial state $\rho^\init_\bet$ is
$\rho_\bet(t,\th)\in\Cbfnn$ which evolves according to,
\beq[eq:rhobet]
i\hbar\dot\rho_\bet = [H(t,\th),\rho_\bet],\
\rho_\bet(0,\th)=\rho^\init_\bet
\eeq
Equivalently,
\beq[eq:rho1]
\rho_\bet(t,\th) = U(t,\th) \rho^\init_\bet U(t,\th)^*
\eeq
where $U(t,\th)\in\Cbfnn$ is the unitary propagator associated with
$H(t,\th)$ which satisfies,
\beq[eq:prop]
i\hbar\dot U = H(t,\th) U,\ U(0,\th)=I_n
\eeq
At each of $\nsa$ sample times in a time interval of duration $\tf$,
measurements are recorded from identical repeated
experiments. Specifically, let $\set{t_\tau}{\tau=1,\ldots,\nsa}$
denote the sample times relative to the start of each experiment. Let
$n_{\alf\bet\tau}$ be the number of times the outcome $\alf$ is
recorded at $t_\tau$ with initial state $\rho^\init_\bet$ from
$\ell_{\bet\tau}$ experiments.  The data set thus consists of all the
outcome counts,
\beq[eq:data th]
D = \set{n_{\alf\bet\tau}}
{\alf=1,\ldots,\nout,\
\bet=1,\ldots,\nin,\
\tau=1,\ldots,\nsa}
\eeq
The {\em configurations} previously enumerated and labeled by
$\gam=1,\ldots,\nconf$ are in this case all the combinations of input
states $\rho^\init_\bet$ and sample times $\tau$, thus $\nconf=\nin\nsa$.
For the POVM $M_\alf$, the model outcome probability per configuration
pair $(\rho^\init_\bet,\ t_\tau)$ is,
\beq[eq:palfbettau]
\bea{rcl}
p_{\alf\bet\tau}(\th)
&=&
\trace\ M_\alf \rho_\bet(t_\tau,\th)
=
\trace\ O_{\alf\tau}(\th) \rho^\init_\bet
\\
O_{\alf\tau}(\th)
&=&
U(t_\tau,\th)^* M_\alf U(t_\tau,\th)
\eea
\eeq
The Maximum Likelihood estimate, $\thml\in\Rbf^{\nth}$, is obtained
as the solution to the optimization problem:
\beq[eq:maxlike th]
\bea{ll}
\mbox{minimize} 
& 
L(D,\th)
=
-\sum_{\alf,\bet,\tau}
n_{\alf\bet\tau}
\log\trace\ O_{\alf\tau}(\th)\rho^\init_\bet
\\
\mbox{subject to} 
& 
\th \in \Th
\eea
\eeq
where $\Th$ is a set of constraints on $\th$. For example, it may be
known that $\th$ is restricted to a region near a nominal value, \eg,
$\Th=\set{\th}{\norm{\th-\th_{\rm nom}} \leq \del}$.  Although this
latter set is convex, unfortunately, the likelihood function,
$L(D,\th)$, is not guaranteed to be convex in $\th$. It is possible,
however, that it is convex in the restricted region $\Th$, for
example, if $\del$ is sufficiently small.

\subsection{Experiment design for Hamiltonian parameter estimation}
\label{sec:ham oed}

Despite the fact that Hamiltonian parameter estimation is not convex,
the (relaxed) experiment design problem is convex. A direct
application of the \crao bound to the likelihood function in
\refeq{maxlike th} results in the following.

\bquote
{\bf Hamiltonian parameter estimation variance lower bound}
Suppose the system generating the data is in the model set used for
estimation, \ie, \refeq{s in m} holds.  For $\ell=[\ell_1 \cdots
\ell_{\nconf}]$ experiments per configuration $(\rho^\init_\bet,t_\tau)$,
suppose $\thh(\ell)\in\Rbf^\nth$ is an unbiased estimate of
$\thtrue\in\Rbf^\nth$.  Under these conditions, the estimation error
variance satisfies,
\beq[eq:var th]
\avg\ \norm{\thh(\ell)-\thtrue}^2
\geq
V(\ell,\thtrue)
=
\trace\
G(\ell,\thtrue)^{-1}
\eeq
where
\beq[eq:var th 1]
\bea{rcl}
\ds
G(\ell,\thtrue) 
&=& 
\ds
\sum_{\bet,\tau}\ \ell_{\bet\tau} G_{\bet\tau}(\thtrue)
\in\Rbf^{\nth \times \nth}
\\&&\\
\ds
G_{\bet\tau}(\thtrue)
&=&
\ds
\left.
\sumalf\
\left(
\frac{
\left(\nabla_\th\ p_{\alf\bet\tau}(\th) \right)
\left(\nabla_\th\ p_{\alf\bet\tau}(\th) \right)^T
}
{p_{\alf\bet\tau}(\th)}
-
\nabla_{\th\th}\ p_{\alf\bet\tau}(\th) 
\right)
\right|_{\th=\thtrue}
\in\Rbf^{\nth\times\nth}
\eea
\eeq
\equote
The relaxed experiment design problem with respect to the surrogate
$\thh$ for $\thtrue$ is,
\beq[eq:expdes th]
\bea{ll}
\mbox{minimize}
&
V(\lam,\thh)
=
\trace\left( 
\sum_{\bet,\tau}\ \lam_{\bet\tau} G_{\bet\tau}(\thh)
\right)^{-1}
\\
\mbox{subject to}
&
\sum_{\bet,\tau}\ \lam_{\bet\tau} = 1
\\
&
\lam_{\bet\tau} \geq 0,\;
\forall\ {\bet,\ \tau}
\eea
\eeq
with optimization variables $\lam_{\bet\tau}$, the distribution of
experiments per configuration $(\rho^\init_\bet,\ t_\tau)$.  The
difference between this and the previous formulation is that there are
no equality constraints on the parameters.  The gradient $\nabla_\th\
p_{\alf\bet\tau}(\th)$ and Jacobian $\nabla_{\th\th}\
p_{\alf\bet\tau}(\th)$ are dependent on the parametric structure of
the Hamiltonian $H(t,\th)$.

\subsection{Example: experiment design for Hamiltonian parameter estimation}
\label{sec:ex hampar}

\begin{figure}[h]
\epsfysize=3.5in
\epsfxsize=6in
\epsfboxit{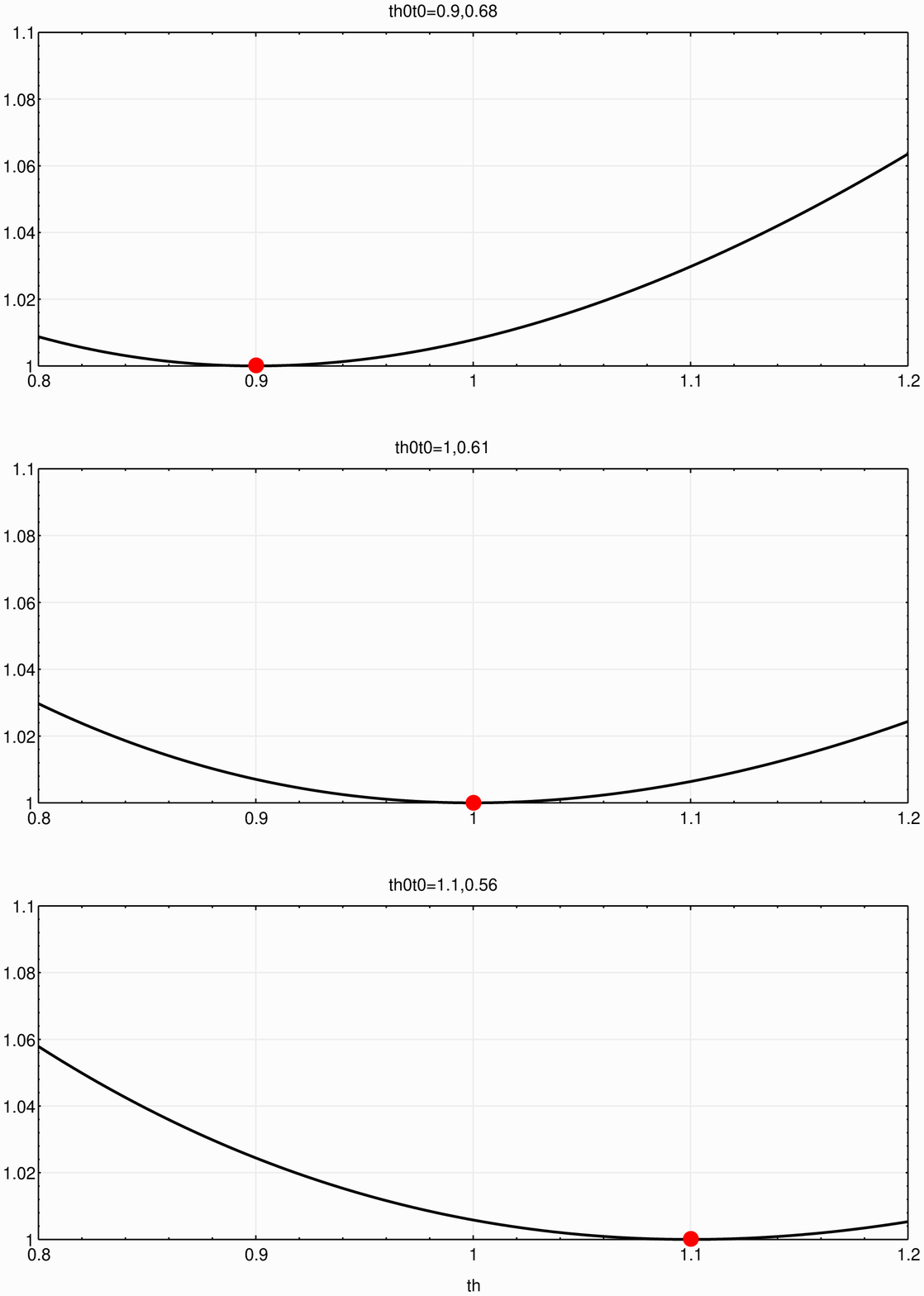}
\caption{ 
Normalized average likelihood function $\avlike(\th)$ with control
$\eps=1$ and input state $\psi^\init=\ket{0}$ for the
true parameter values and associated optimal recording times as
indicated above and given in \refeq{tables e1}.
}
\label{fig:avlike_e1r1}
\end{figure}

Consider the system Hamiltonian,
\beq
H = \th^\true {\eps}\ 
\left(X + Z \right)/\sqrt{2},
\eeq
with {\em constant} control $\eps$. The goal is to select the control
to make the Hadamard logic gate, $\Uhad=(X+Z)/\sqrt{2}$. If $\th^\true$
were known, then the control $\eps=1/\th^\true$ would produce the
Hadamard (to within a scalar phase) at time $t=\pi/2$, that is,
\beq
U(t=\pi/2)
=
\exp\seq{-i(\pi/2)H(\eps=1/\th^\true)}
=
-i\ \Uhad
\eeq
We assume that only the estimate $\thh$ of $\th^\true$ is available.
Using the estimate and knowledge of the Hamiltonian model structure,
the control is $\eps=1/\thh$. This yields the {\em actual} gate at
$t=\pi/2$,
\beq
\bea{rcl}
U_{\rm act}
&=& 
\exp\seq{-i(\pi/2)H(\eps=1/\thh)}
=
-i\ \Uhad\ \exp\seq{-i \del\ (\pi/2)\ \Uhad}
\\
\del &=& \th^\true/\thh-1
\eea
\eeq
Since the parameter estimate, $\thh$, is a random variable, so is the
normalized parameter error $\del$.  Assuming the estimate is unbiased,
the expected value of the worst-case gate fidelity \refeq{wc fidelity}
is given explicitly by,
\beq[eq:wcfid]
\avg\
\min_{\norm{\psi}=1}\
\left|
\left(\Uhad \psi\right)^*
\left(U_{\rm act}\psi\right)
\right|^2
=
\avg\ \cos^2\left(\frac{\pi}{2}\ \del\right)
\approx
1-\left(\frac{\pi}{2}\right)^2\ \avg(\del^2)
\eeq
Consider the case where the system is in the model set, the POVMs are
projectors in the computational basis $(\ket{0},\ \ket{1})$, and the
configurations consist of combinations of input states and sample
times. Specifically, the example problem is as follows:
\beq
\bea{ll}
\mbox{{model Hamiltonian}}
&
H(\eps,\ \th) = 
\th\ \eps
\left( X + Z \right)/\sqrt{2}
\\&\\
\mbox{{true Hamiltonian}}
&
H^\true=\thtrue\eps\ \left( X + Z \right)/\sqrt{2}
\\&\\
\mbox{{POVM}}
&
M_1 = \ket{0}\bra{0},
\;\;
M_2 = \ket{1}\bra{1}
\\&\\
\mbox{{configurations}}
&
\left\{
\bea{c}
\mbox{sample times}
\\
\\
t_k=\del(k-1),\ k=1,\ldots,100,\ \del=(\pi/2)/99
\\
\\
\mbox{with pure input state}
\\
\\
\psi^\init=\ket{0}\;
\mbox{or}\;
\psi^\init=\Uhad\ket{0}
\eea
\right.
\eea
\eeq
In this example, with a single parameter and a single input state, the
optimal experiment design problem \refeq{expdes th} becomes:
\beq[eq:expdes single th]
\bea{ll}
\mbox{minimize}
&
V(\lam,\th^\surr)
=
\left(
\sum_{\tau}\ \lam_{\tau} g_{\tau}(\th^\surr)
\right)^{-1}
\\
\mbox{subject to}
&
\sum_{\tau}\ \lam_{\tau} = 1
\\
&
\lam_{\tau} \geq 0,\;
\forall\ {\tau}
\eea
\eeq
The trace operation in \refeq{expdes th} is eliminated because the
matrix $G_{\bet\tau}(\th^\surr)$ is now the scalar,
\beq[eq:g sinlge th]
g_{\tau}(\th^\surr)
=
\left.
\sumalf\
\left(
\frac{
\left(\nabla_\th\ p_{\alf\tau}(\th) \right)
\left(\nabla_\th\ p_{\alf\tau}(\th) \right)^T
}
{p_{\alf\tau}(\th)}
-
\nabla_{\th\th}\ p_{\alf\tau}(\th) 
\right)
\right|_{\th=\th^\surr}
\in\Rbf
\eeq
The solution can be determined directly: concentrate all the
experiments at the recording time $t_\tau$ where $g_{\tau}(\th^\surr)$ is a
maximum, specifically,
\beq[eq:topt]
t^\opt = \set{t_s}{g_{s}(\th^\surr) \geq g_{\tau}(\th^\surr),\ \forall s,\tau}
\eeq 
The following tables show the minimum number of experiments at the
optimal recording time, $t^\opt$, in order to achieve 0.01 accuracy
(deviation) in the parameter estimate with $\th^\surr=\th^\true$.  The
two cases shown are for the two input states with the control set to
unity.
\beq[eq:tables e1]
\bea{cc}
\bea{c}
\eps = 1
\\
\psi^\init=\ket{0}
\\
\\
\renewcommand{\arraystretch}{1}
\bea{|c||c|c|}
\hline
\th^\surr=\th^\true & t^\opt/(\pi/2) & \nex\\
\hline\hline
0.9 & 0.68 & 8,876\\
1.0 & 0.61 & 10,957\\
1.1 & 0.56 & 13,262\\
\hline
\eea
\eea
&
\bea{c}
\eps = 1
\\
\psi^\init=\Uhad\ket{0}
\\
\\
\renewcommand{\arraystretch}{1}
\bea{|c||c|c|}
\hline
\th^\surr=\th^\true & t^\opt/(\pi/2) & \nex\\
\hline\hline
0.9 & 1.0 & 2052\\
1.0 & 1.0 & 2069\\
1.1 & 1.0 & 2052\\
\hline
\eea
\eea
\eea
\eeq
To make the Hadamard the control update is $\eps=1/\thh$ which follows
from the (risky) assumption that the estimate, $\thh$, is perfectly
correct.  For any of the true values from the above table
\refeq{tables e1}, all the estimates have the same accuracy. Hence,
the average value of the worst-case gate fidelity after the update is
approximately,
\beq
\avg\ 
\min_{\norm{\psi}=1}
\left|
\left(\Udes \psi\right)^*
\left( U_{\rm act}\psi \right)
\right|^2
\approx
1-\left(\frac{\pi}{2}\right)^2\ (0.01^2)
=
1-0.00024672
=
0.999753
\eeq
The ease of obtaining the estimate by minimizing the negative
log-likelihood function can be determined by examining the {\em
average likelihood function},
\beq[eq:avlike]
\avlike(\th)
=
\sum_{\alf,\gam}
\ell_\gam\
p_{\alf\gam}(\th^\true)\ 
\log p_{\alf\gam}(\th)
\eeq
which is obtained from \refeq{maxlike th} and \refeq{av nalfgam}.
Figure \ref{fig:avlike_e1r1} shows plots of normalized\footnote{
The plots show $\avlike(\th)$ divided by its minimum value, thus
{\em normalized} to have a minimum of unity.
}
$\avlike(\th),\ 0.8 \leq \th \leq 1.2$ for the three true parameter
values and corresponding optimal recording times from the table in
\refeq{tables e1} with control $\eps=1$ and initial state $\ket{0}
$. In all three cases, over the $\th$ range shown, $\avlike(\th)$ is
convex.

\begin{figure}[h]
\epsfysize=3.5in
\epsfxsize=6in
\epsfboxit{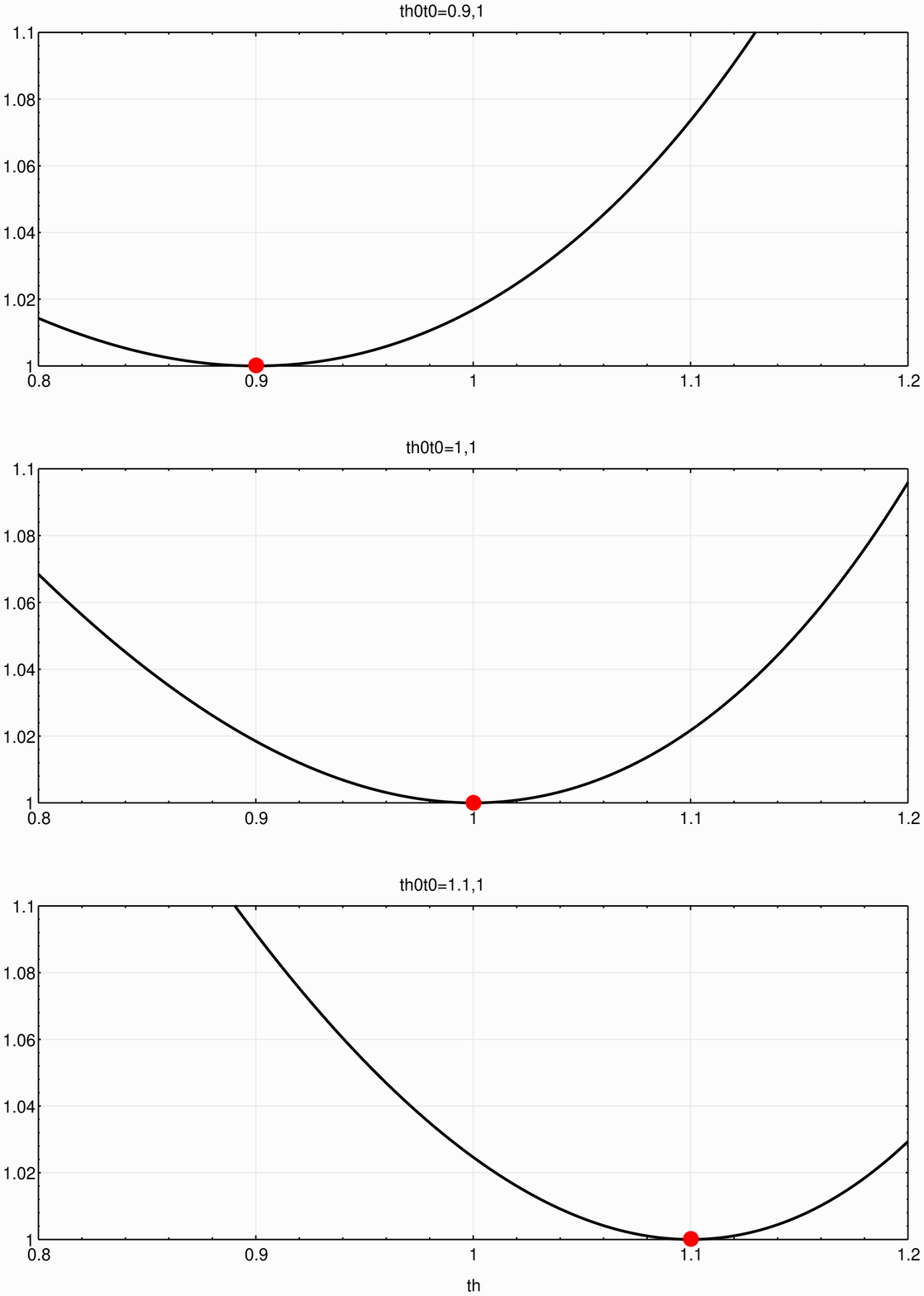}
\caption{
Normalized average likelihood function $\avlike(\th)$ with control
$\eps=1$ and initial state $\psi^\init=\Uhad\ket{0}$
for the true parameter values and associated optimal recording times
as indicated above and given in \refeq{tables e1}.
}
\label{fig:avlike_e1rh1}
\end{figure}

Figure \ref{fig:avlike_e1rh1} shows plots of normalized
$\avlike(\th),\ 0.8 \leq \th \leq 1.2$ for the three true parameter
values and corresponding optimal recording times from the table in
\refeq{tables e1} with control $\eps=1$ and initial state
$\Uhad\ket{0}$.  For initial state $\ket{0}$, a range of 8876 to 13262
experiments are needed at the optimal recording time to achieve 0.01
deviation in the estimate. With the initial states $\Uhad\ket{0}$ the
same accuracy only requires about 2000 experiments. This difference
can be inferred partly by comparing the curvature in the plots in
Figure \ref{fig:avlike_e1r1} with \ref{fig:avlike_e1rh1}; as they are
plotted on the same normalized scale. Note the increased curvature of
$\avlike(\th)$ in Figure \ref{fig:avlike_e1rh1} in the neighborhood of
the true value.

\begin{figure}[h]
\epsfysize=3.5in
\epsfxsize=6in
\epsfboxit{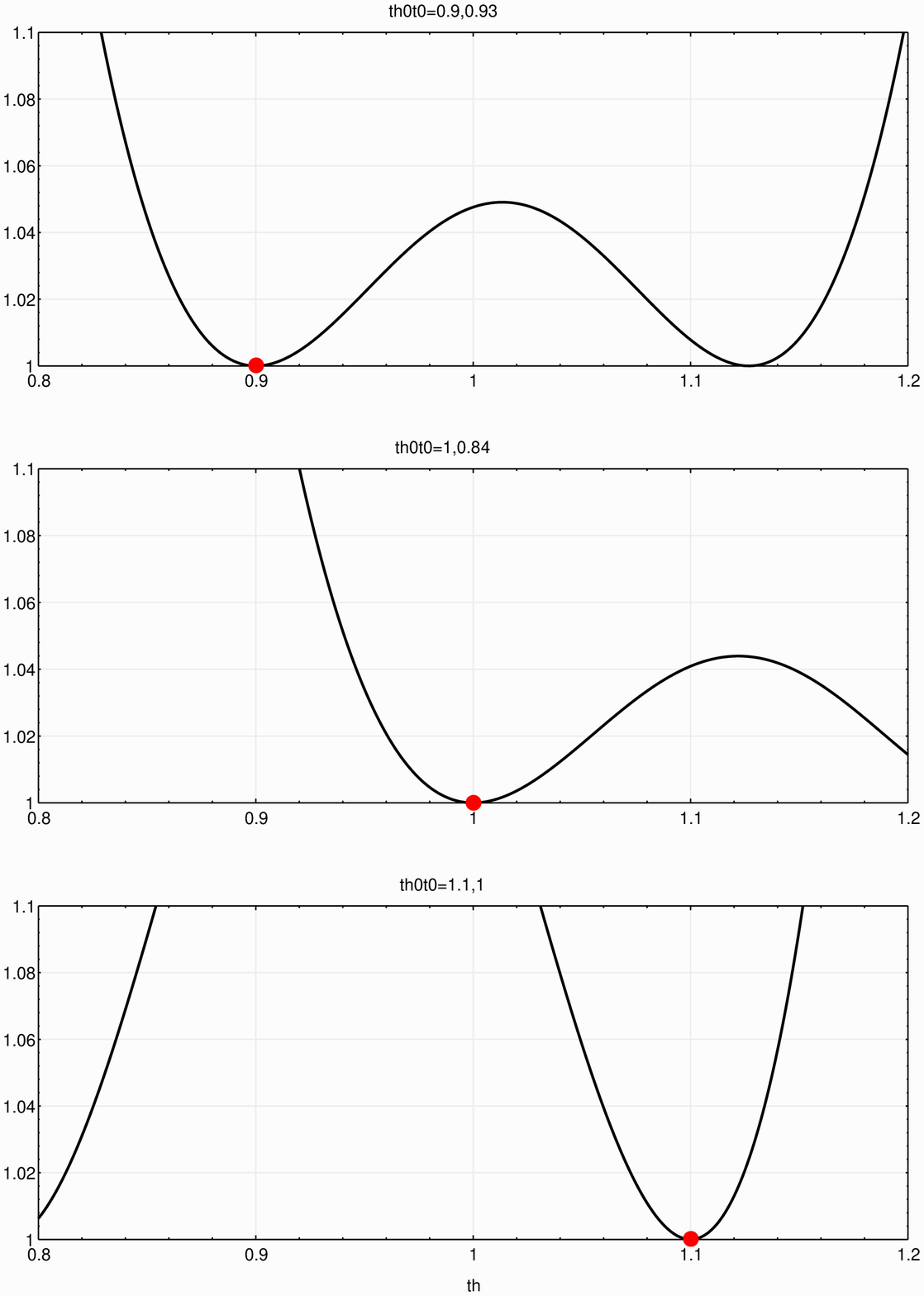}
\caption{
Normalized average likelihood function $\avlike(\th)$ with control
$\eps=5$ and initial state $\psi^\init=\ket{0}$
for the true parameter values and associated optimal recording times
as indicated above and given in \refeq{table e5}.
}
\label{fig:avlike_e5r1}
\end{figure}

If we further increase the control effort, say to $\eps=5$,
the number of experiments required to achieve 0.01 accuracy
is significantly reduced as seen in the following table.
\beq[eq:table e5]
\bea{c}
\eps = 5
\\
\psi^\init=\ket{0}
\\
\\
\renewcommand{\arraystretch}{1}
\bea{|c||c|c|}
\hline
\th^\surr=\th^\true & t^\opt/(\pi/2) & \nex\\
\hline\hline
0.9 & 0.93 & 98\\
1.0 & 0.84 & 121\\
1.1 & 1.00 & 122\\
\hline
\eea
\eea
\eeq
However, Figure \ref{fig:avlike_e5r1} shows clearly that the average
likelihood function is now significantly more oscillatory, and
certainly not convex over the range shown. It is of course convex in a
much smaller neighborhood of the true value. This would require,
therefore, very precise prior knowledge about the true system.  Thus
we see a clear tradeoff between the number of experiments to achieve a
desired estimation accuracy and the ease of obtaining the estimate as
seen by the convexity, or lack thereof, with respect to minimizing the
likelihood function, which is the optimization objective.


\psfrag{th}{$\bea{c}{}\\ \th\eea$}
\psfrag{th0t0=0.9,0.68}{$\thtrue=0.9,\ t^\opt/(\pi/2)=0.68$}
\psfrag{th0t0=1,0.61}{$\thtrue=1,\ t^\opt/(\pi/2)=0.61$}
\psfrag{th0t0=1.1,0.56}{$\thtrue=1.1,\ t^\opt/(\pi/2)=0.56$}

\psfrag{th}{$\bea{c}{}\\ \th\eea$}
\psfrag{th0t0=0.9,1}{$\thtrue=0.9,\ t^\opt/(\pi/2)=1$}
\psfrag{th0t0=1,1}{$\thtrue=1,\ t^\opt/(\pi/2)=1$}
\psfrag{th0t0=1.1,1}{$\thtrue=1.1,\ t^\opt/(\pi/2)=1$}

\psfrag{th}{$\bea{c}{}\\ \th\eea$}
\psfrag{th0t0=0.9,0.93}{$\thtrue=0.9,\ t^\opt/(\pi/2)=0.93$}
\psfrag{th0t0=1,0.84}{$\thtrue=1,\ t^\opt/(\pi/2)=0.84$}
\psfrag{th0t0=1.1,1}{$\thtrue=1.1,\ t^\opt/(\pi/2)=1$}

\clearpage
\section{Summarizing Maximum Likelihood Estimation 
\& Optimal Experiment Design}
\label{sec:summary}

The results presented show that an efficient numerical method based on
convex programming is possible for optimizing the experiment for state
and process tomography. In addition, the estimation of the state
and/or process using data from non-continuing measurements is
copacetic with Maximum Likelihood Estimation. Both the experiment
design and estimation work naturally together and both can be solved
using convex optimization methods.

\subsection*{Maximum likelihood estimation}

The general form for estimating the parameter $\pit$ is obtained as
the solution to the optimization problem,
\beq[eq:mle gen]
\mathbox{
\bea{ll}
\mbox{minimize}
&
L({\pi}) = -\sum_{\alf,\gam}\ 
n_{\alf\gam}\ \log\ p_{\alf\gam}({\pi})
\\&\\
\mbox{subject to}
&
{\pi} \in \Pi 
\eea
}
\eeq
Under the assumption that the system generating the data is in the
model set used for estimation, the estimate, $\pi^{\rm ML}$, the
solution to \refeq{mle gen}, is unbiased and has the asymptotic
variance,
\beq[eq:varmle]
\bea{c}
\avg\ \norm{\pi^{\rm ML}-\pit}^2
\to
\frac{1}{\nex}\ V(\lam,\pit)
\;\;\;
\mbox{as $\nex\to\infty$}
\eea
\eeq
where $\lam$ is the vector of fraction of experiments per
configuration, and $V(\lam,\pit)$ is obtained from the \crao
inequality.

\subsection*{Optimal experiment design}

The general form for estimating the configuration distribution $\lam$
is obtained as the solution to the optimization problem,
\beq
\mathbox{
\bea{ll}
\mbox{minimize}
&
V({\lam},\pih) = 
\trace\ 
\left(
\sumgam\ {\lam}_\gam\ G_\gam(\pih)
\right)^{-1}
\\&\\
\mbox{subject to}
&
\sumgam\ {\lam}_\gam=1,
\;\;\;
{\lam}_\gam \geq 0
\eea
}
\eeq
where $\pih$ is a surrogate for $\pit$.

Tables \ref{tab:mle} and \ref{tab:oed} summarize the class of
estimation and experiment design problems, respectively.

\newpage

\begin{small}
\begin{table}[t]
\btab{|c||c|c|c|}
\hline
&&&
\\
\blue{objective} & \blue{$p_{\alf\gam}(\pi)$} & \blue{$\pi\in\Pi$} 
& \blue{comment}
\\
&&&
\\
\hline\hline
\black{\btab{l}
Hamiltonian\\
parameter\\
estimation
\etab}
& $p_{\alf\gam}(\red{\th})=\trace\ O_{\alf\gam}(\red{\th})\rho_\gam$ 
& $\norm{\red{\th}-\th_{\rm nom}} \leq \del$
& 
\btab{l}
not convex\\ 
in $\red{\th}$\\
many local\\ 
minima
\etab
\\
\hline
\black{\btab{l}
state\\
estimation
\etab}
&
$p_{\alf\gam}(\red{\rho}) = \trace\ O_{\alf\gam}\ \red{\rho}$
& 
$\trace\ \red{\rho}=1,\; \red{\rho} \geq 0$ 
& convex in  $\red{\rho}$
\\
\hline
\black{\btab{l}
state\\
distribution\\
estimation
\etab}
&
$
\bea{l}
p_{\alf\gam}(\red{f})=a_{\alf\gam}^T \red{f}
\\
(a_{\alf\gam})_i = \trace\ O_{\alf\gam}\ \rho_i
\eea
$
& $\sum_i \red{f_i}=1, \;\;\; \red{f_i}\geq 0$
& convex in $\red{f}$
\\
\hline
\black{\btab{l}
OSR
\\
fixed basis
\\
($B_i$)
\etab}
&
$
\bea{l}
p_{\alf\gam}(\red{X}) = \trace\ R_{\alf\gam}\ \red{X}
\\
\left[ R_{\alf\gam} \right]_{ij} = \trace\ B_j^* M_{\alf\gam} B_i\ \rho_\gam
\eea
$
&
$
\bea{l}
\red{X} \geq 0
\\
\sum_{i,j}\ \red{X_{ij}}\ B_i^*\ B_j = I
\eea
$
& convex in $\red{X}$
\\
\hline
\black{\btab{l}
OSR
\\distribution
\\
($\Kb_i$ basis)
\etab}
&
$
\bea{l}
p_{\alf\gam}(\red{q}) 
= 
a_{\alf\gam}^T \red{q} 
\\
(a_{\alf\gam})_i = \trace\ M_{\alf\gam} \Kb_i \rho_\gam \Kb_i^*
\eea
$
&
$\sum_i \red{q_i}=1, \;\;\; \red{q_i} \geq 0$
& convex in $\red{q}$
\\
\hline
\etab
\caption{
Summary of maximum likelihood estimation.
Except for Hamiltonian parameter estimation, 
all other cases are convex optimization problems.
}
\label{tab:mle}
\end{table}
\end{small}
%

\begin{small}
\begin{table}[b]
\btab{|c||c|c|}
\hline
&&
\\
\blue{objective} & \blue{$p_{\alf\gam}(\pi^\surr)$} 
& 
\blue{$G_\gam(\pi^\surr)$} 
\\
&&
\\
\hline\hline
\black{\btab{l}
Hamiltonian\\
parameter\\
estimation
\etab}
& $p_{\alf\gam}(\blues{\th})=\trace\ O_{\alf\gam}(\blues{\th})\rho_\gam$ 
& 
$
\sumalf\left(
\fracds{1}{p_{\alf\gam}}
(\nabla_\th\ p_{\alf\gam})(\nabla_\th\ p_{\alf\gam})^T
-\nabla_{\th\th}\ p_{\alf\gam}
\right)
$
\\
\hline
\black{\btab{l}
state\\
estimation
\etab}
&
$p_{\alf\gam}(\blues{\rho}) = \trace\ O_{\alf\gam}\ \blues{\rho}$
&
$ 
\Ceq^*\left(
\sumalf \fracds{1}{p_{\alf\gam}} 
(\vec\ O_{\alf\gam}) (\vec\ O_{\alf\gam})^*
\right)\Ceq
$
\\
\hline
\black{\btab{l}
state\\
distribution\\
estimation
\etab}
&
$
\bea{l}
p_{\alf\gam}(\blues{f}) = a_{\alf\gam}^T \blues{f}
\\
(a_{\alf\gam})_i = \trace\ O_{\alf\gam} \rho_i
\eea
$
& 
$
\Ceq^*\left(
\sumalf \fracds{1}{p_{\alf\gam}} a_{\alf\gam} a_{\alf\gam}^T
\right)\Ceq
$
\\
\hline
\black{\btab{l}
OSR
\\
fixed basis
\\
($B_i$)
\etab}
&
$
\bea{l}
p_{\alf\gam}(\blues{X}) = \trace\ R_{\alf\gam}\ \blues{X}
\\
\left[ R_{\alf\gam} \right]_{ij} = \trace\ B_j^* M_{\alf\gam} B_i\ \rho_\gam
\eea
$
&
$
\Ceq^*\left(
\sumalf \fracds{1}{p_{\alf\gam}} 
(\vec\ R_{\alf\gam}) (\vec\ R_{\alf\gam})^*
\right)\Ceq
$
\\
\hline
\black{\btab{l}
OSR
\\distribution
\\
($\Kb_i$ basis)
\etab}
&
$
\bea{l}
p_{\alf\gam}(\blues{q}) 
= 
a_{\alf\gam}^T \blues{q} 
\\
(a_{\alf\gam})_i = \trace\ M_{\alf\gam} \Kb_i \rho_\gam \Kb_i^*
\eea
$
&
$
\Ceq^*\left(
\sumalf \fracds{1}{p_{\alf\gam}} a_{\alf\gam} a_{\alf\gam}^T
\right)\Ceq
$
\\
\hline
\etab
\caption[oed]{ Summary of optimal experiment designs. These are convex
optimization problems in all cases. The matrix $\Ceq$ comes from the
associated parameter equality constraints, \eg, 
$\trace\ \blues{\rho} =1,\
\sum_i \blues{f_i}=1$, $\sum_{i,j}\ \blues{X_{ij}}\ B_i^*\ B_j = I$, \etc.  }
\label{tab:oed}
\end{table}
\end{small}

\clearpage
\section{Iterative Adaptive Control}
\label{sec:iter adapt}

Quantum process tomography can be used for adaptive control design as
described in \cite{KosutRW:03}. Adaptive control systems are in
general one of two types \cite{AstromW:95,AndersonETAL:86}: (1) {\em
indirect adaptive control} -- use the data to first determine
parameters in a system model, then based on the model determine the
control parameters, and (2) {\em direct adaptive control} -- use the
data to directly select control parameters by comparing actual
performance to an ideal.  Applications of direct adaptive control (and
learning principles) to quantum systems can be found in
\cite{PhanR:97,ZhuR:98,PhanR:99}.  It could be argued that the direct
approach is simpler in that a system model is not needed.
However, this is a little deceptive because in effect the {\em
closed-loop system} is being modeled indirectly via the ideal
performance. This is made more explicit in the {\em
unfalsification/invalidation} approach to adaptive control,
\cite{SafonovT:97}, \cite{Kosut:01}.

\subsection{Indirect adaptive control}
\label{sec:indirect adapt}

Hamiltonian parameter estimation and associated optimal experiment
design can be combined in an iterative indirect adaptive control
approach as depicted in Figure \ref{fig:adapt}.

\begin{small}
\begin{figure}[h]
\unitlength 1.2mm
\thicklines
\centering
\begin{picture}(104,75)(-10,-15)
\put(0,0){\framebox(30,15){\fs Controller}}
\put(30,7.5){\vector(1,0){20}}
\put(50,0){\framebox(30,15){\fs System}}
\put(80,7.5){\line(1,0){10}}
\put(0,25){\framebox(30,15)
{\shortstack{\fs Control\\ \fs Design}}}
\put(50,25){\framebox(30,15)
{\shortstack{\fs Model\\ \fs Parameter\\ \fs Estimator}}}
\put(-20,11){\vector(1,0){20}}
\put(-23,12){$r$}
\put(-10,4){\vector(1,0){10}}
\put(-10,4){\line(0,-1){14}}
\put(-10,-10){\line(1,0){100}}
\put(90,-10){\line(0,1){42.5}}
\put(90,32.5){\vector(-1,0){10}}
\put(40,7.5){\line(0,1){25}}
\put(40,32.5){\vector(1,0){10}}
\put(65,40){\line(0,1){10}}
\put(15,50){\vector(0,-1){10}}
\put(15,25){\vector(0,-1){10}}
\put(15,50){\line(1,0){10}}
\put(25,42.5){\framebox(30,15)
{wait}}
\put(65,50){\vector(-1,0){10}}
\put(70,50){$\th_\itip$}
\put(8,50){$\thh_\iti$}
\put(40,3){$\eps_\iti$}
\put(93,7.5){$n_\iti$}
\end{picture}
\caption{Iterative adaptive control}
\label{fig:adapt}
\end{figure}
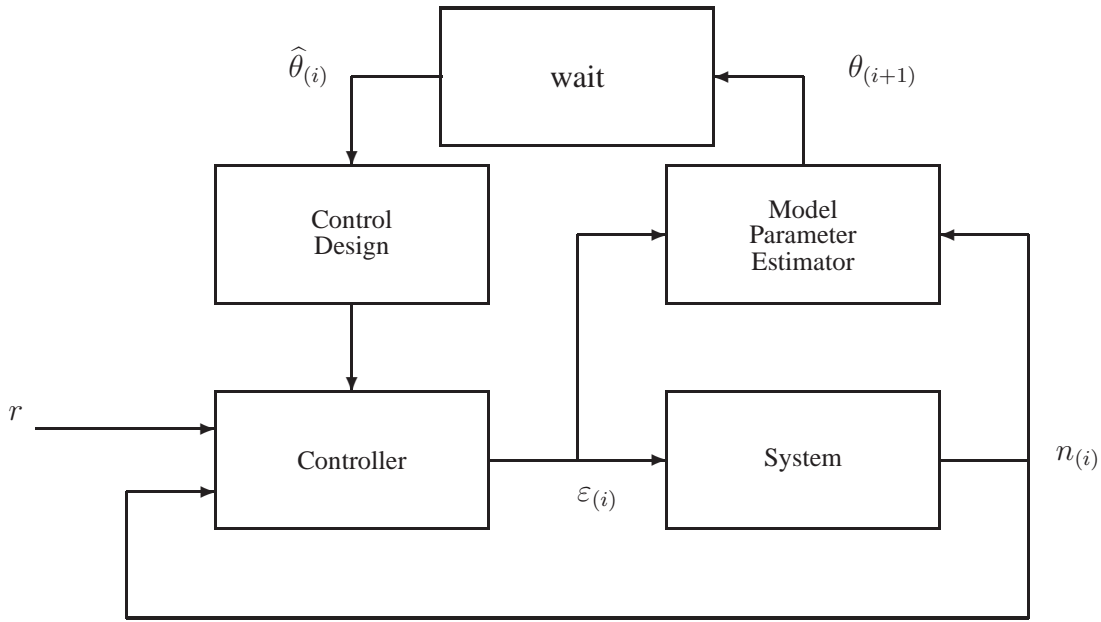
\end{small}

Typical steps in the iteration are:
\beq[eq:indirect adapt]
\bea{ll}
\mbox{control design}
&
\ds
\eps^\iti
=
\arg\opt_{\eps}\ J(\eps,\th^\iti)
\\&\\
\mbox{experiment design}
&
\ds
\ell^\iti = {\bf round}(\nex\ \lam^\iti)
\\
&
\ds
\lam^\iti
= 
\arg\min
\sett{
V(\lam,\eps^\iti,\th^\iti)
}{
\lam\geq 0,\ \sum_\gam\ \lam_\gam=1
}
\\&\\
\mbox{collect data}
&
\ds
D^\iti =  \sett{ n_{\alf\gam}^\iti }{ 
\alf=1,\ldots,\nout,\ \gam=1,\ldots,\nconf
}
\\&\\
\mbox{estimate parameters}
&
\ds
\th^\itip
=
\arg\min
\sett{ L(D^\iti,\th) }{ \norm{\th-\th_{\rm nom}} \leq \del }
\eea
\eeq
The ``opt'' in the control design step could be to maximize worst-case
gate fidelity \refeq{wc fidelity},
\beq[eq:max wc fid]
\eps^\iti
=
\max_\eps\
\underbrace{
\min_{\norm{\psi}=1}
\left|
\left(\Udes \psi\right)^*
\left( U(\tf,\eps,\th^\iti)\psi \right)
\right|^2
}_{\ds J(\eps,\th^\iti)}
\eeq
where $U(t,\eps,\th)$ is the propagator arising from the parametric
Hamiltonian model. The control design step is not necessarily convex.
Even if it were, it is not yet known under what conditions the
complete iterative procedure will converge to the optimal control, or
converge at all \cite{KosutRW:03}. For example, in the simulations to
follow convergence to the optimum is dependent on the initial
parametrization. The properties of this type of iteration remain an
area for further study.

\subsubsection*{Example}

The spin-coherent photon transmitter/receiver system proposed in
\cite{VriETAL:00,YabETAL:03} creates quantum logic gates by
manipulating electron spin via external potentials (gate voltages) to
effect the {\em g-factors} in the semiconductor material in the
presence of an external (rotating) magnetic field (\cite{Awsc:03}
explores {\em g-tensor} control without the rotating field). Following
\cite[III,Ch.12-9]{FeynmanLS:65} on models of spin systems, an
idealized model of the normalized Hamiltonian in the rotating frame of
a two-qubit gate under ``linear g-factor control'' is given by,
\[
\bea{rcl}
H &=& H_1 + H_2 +H_{12}
\\&&\\
H_1 
&=& 
\frac{1}{2} 
\left[ 
{\eps_{1z}}
\om_0(Z\otimes I_2)
+
{\eps_{1x}}
\om_1(X\otimes I_2)
\right]
\\&{}&\\ 
H_2
&=&
\frac{1}{2} 
\left[ 
{\eps_{2z}}
\om_0(I_2\otimes Z)
+
{\eps_{2x}}
\om_1(I_2\otimes X)
\right]
\\&&\\ 
H_{12}
&=& 
\ds
{\eps_c} 
\om_c
\left(
X^{\otimes 2}+Y^{\otimes 2}+Z^{\otimes 2}
\right)
\eea
\]
The design goal is to use the 5 controls
$
({
\eps_{1z},\
\eps_{1x},\
\eps_{2z},\
\eps_{2x},\
\eps_c
}
)
$
to make the Bell transform,
\[
\Ubell
=
\frac{1}{\sqrt{2}}
\left[
\bea{cccc}
1 &  0 &  1 &  0 \\
0 &  1 &  0 &  1 \\
0 &  1 &  0 &  -1 \\
1 &  0 &  -1 &  0
\eea
\right]
\]
One of the many possible decompositions of the Bell transform is the
following:
\[
\Ubell
=
(\Uhad \otimes I_2)\
\Usqsw\
( X^{-1/2} \otimes X^{1/2} )
\Usqsw\
( I_2 \otimes X)
\]
Each operation in this sequence uses only the single qubit and swap
``gates'' produced by simultaneously pulsing the 5 controls as shown
in the following table.
\begin{figure}[h]
{
\renewcommand{\arraystretch}{2}
\[
\bea{|cc|cc|c||c||c|}
\hline
\eps_{1z} & \eps_{1x} & \eps_{2z} & \eps_{2x} & \eps_c & \Del t & \mbox{gate}\\
\hline\hline
0&0&0&1&0&\frac{\pi}{{\om_1}}&
-iI_2\otimes X \\
\hline
0&0&0&0&1&\frac{\pi}{8\om_c}&
e^{-i\frac{\pi}{8}} \Usqsw \\
\hline
0&0&0&1&0&\frac{\pi}{2{\om_1}}&
e^{-i\frac{\pi}{4}} I_2\otimes X^{1/2} \\
\hline
0&1&0&0&0&\frac{3\pi}{2{\om_1}}&
e^{-i\frac{3\pi}{4}} X^{-1/2}\otimes I_2 \\
\hline
0&0&0&0&1&\frac{\pi}{8\om_c}&
e^{-i\frac{\pi}{8}} \Usqsw \\
\hline
\frac{\om_\had}{\om_0\sqrt{2}}&\frac{\om_\had}{{\om_1}\sqrt{2}}&
0&0&0&
\frac{\pi}{\om_\had}&-i\Uhad\otimes I_2 \\
\hline
\eea
\]
}
\caption{Pulse control table}
\end{figure}

\begin{figure}[h]
\begin{tabular}{cc}
\epsfysize=5.5in
\epsfxsize=3in
\epsfmaxlike{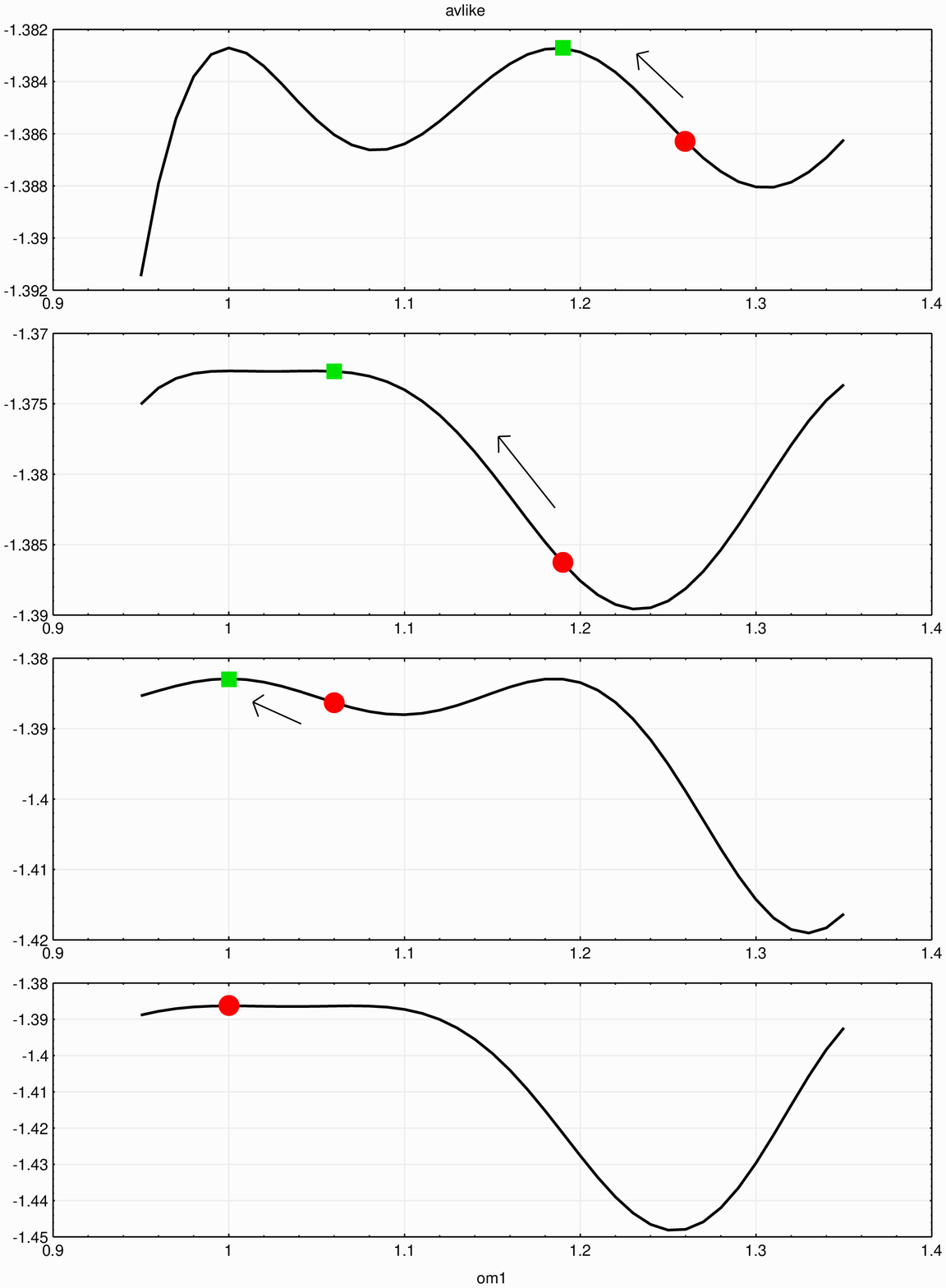}
&
\epsfysize=5.5in
\epsfxsize=3in
\epsfmaxlike{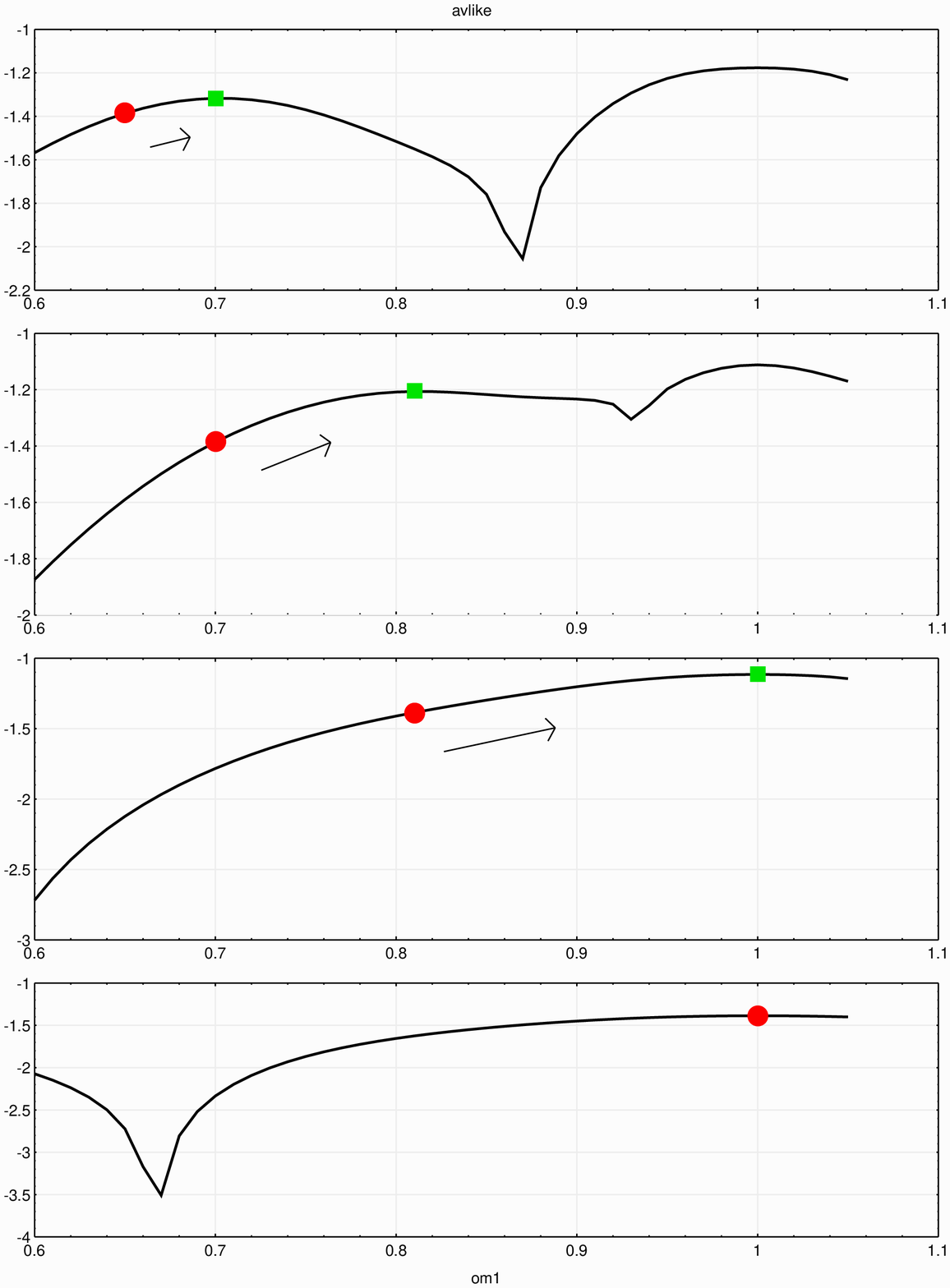}
\end{tabular}
\caption{\small Iterative adaptation for $\nsa=1$ at $\tf(\omh_1)$ for
two starting values of $\omh_1$.}
\label{fig:avlike2cd2eli}
\end{figure}

The resulting gate at the final time, $\tf$, is $\Ubell$ to within a
scalar phase:
\[
U(\tf) 
= 
e^{-i\frac{\pi}{4}}\ \Ubell,
\;\;\;
\tf 
=
\left(
\frac{3}{{\om_1}}+\frac{1}{4\om_c}+\frac{1}{\om_\had}
\right)\pi
\]
Suppose the only unknown parameter is $\om_1$.  Consider the following
simplified version of \refeq{indirect adapt}:
\[
\bea{ll}
\mbox{control design}
&
\eps^\iti
=
\epsb(\omh_1^\iti),
\;\;
\tf^\iti = \tf(\omh_1^\iti)
\\
\mbox{estimation}
&
\omh_1^\itip
=
\ds
\arg\min_{\om_1}\
\avg\ L(\om_1,\ \eps^\iti)
\eea
\]
where the control design function $\epsb(\omh_1^\iti)$ represents the
pulse design from the above table, and where the average likelihood
function follows from the description in Section \refsec{mle hampar}
with the following parameters:
\[
\bea{ll}
\mbox{single initial state}
&
\rho^\init=\ket{0}\bra{0}
\;\;
(\bet=1)
\\
\mbox{POVM}
&
M_1=\ket{0}\bra{0},
\;\;
M_2=\ket{1}\bra{1}
\;\;
(\nout=2)
\\
\mbox{sample times}
&
\mbox{
either $\seq{\tf(\omh_1),\ (\nsa=1)}$
or
$\seq{\tf(\omh_1)/2,\ \tf(\omh_1),\ (\nsa=2)}$
}
\eea
\]
Using Hamiltonian parameters $(\om_0^\true=1,\ \om_1^\true=0.01,\
\om_c^\true=0.01)$, Figure \reffig{avlike2cd2eli} shows $\avg\
L(\omh_1,\ \nsa=1)$ vs. $\omh_1/\om_i^\true$ for sequences of adaptive
iterations using the estimate $\omh_1$ obtained from a local hill
climbing algorithm, \ie, the local maximum of the average likelihood
function is obtained. The estimation is followed by a control using
the estimated value in the pulse control table. In the two cases shown
the algorithm converges to the true value.


\psfrag{avlike}{$\avlike(\omh_1,\rho_0)$}
\psfrag{om1}{$\omh_1/\om_1$}

\begin{figure}[h]
\begin{tabular}{c}
$\norm{U(\tf(\omh_1),\ \epsb(\omh_1))-\Udes}_{\rm frob}$
\\
\epsfysize=4in
\epsfxsize=6.25in
\epsfmaxlike{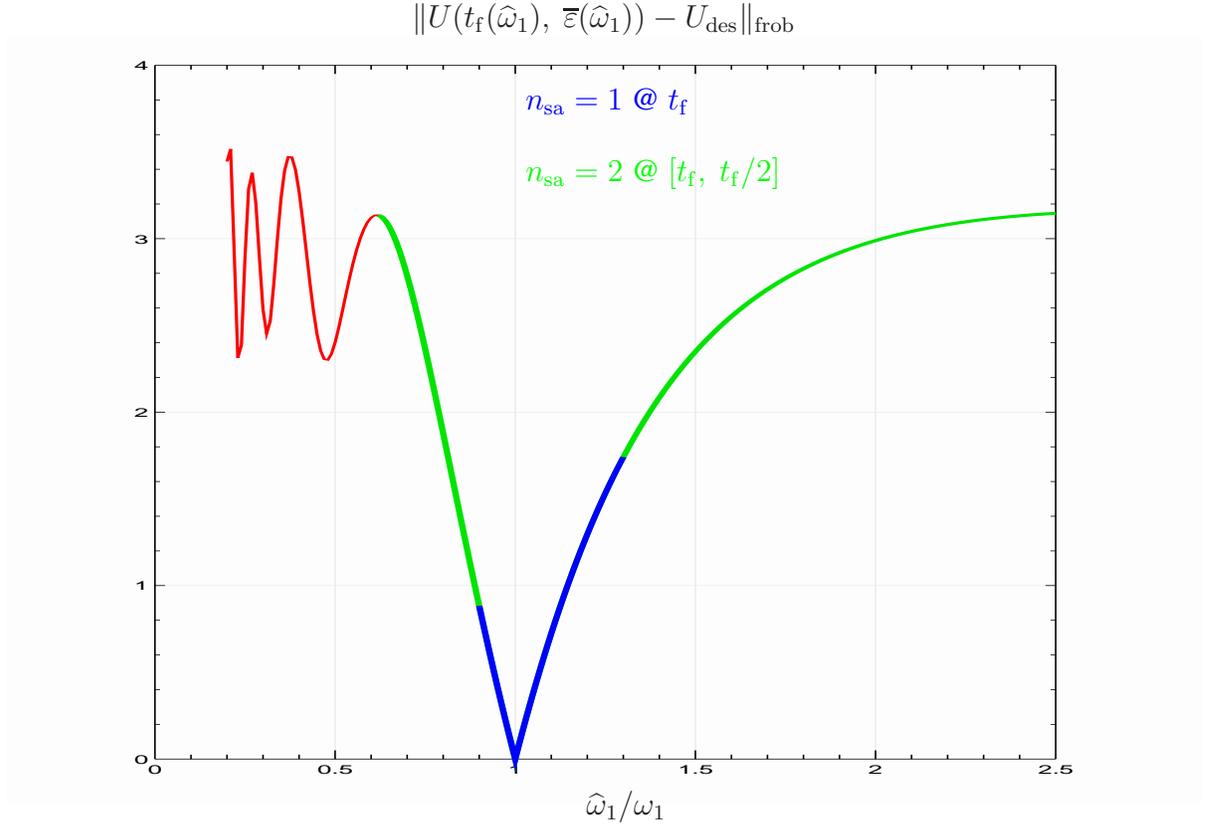}
\end{tabular}
\caption{Regions of convergence.}
\label{fig:directerr01}
\end{figure}

Although not shown, the algorithm does not converge from all initial
values of $\omh_1$.  Figure \reffig{directerr01} shows
$\norm{U(\tf(\omh_1),\ \epsb(\omh_1))-\Udes}_{\rm frob}$ vs.  estimate
$\omh_1/\om_1^\true$ with the control from the table.  The function is
clearly not convex. The region of convergence for $\nsa=1$ and
$\nsa=2$ sample times are shown in blue and green, respectively. The
region of attraction is increased for $\nsa=2$.  These results, of
course, are specific to this example and can not be generalized. To
re-iterate, conditions for convergence, region of attraction, and so
on, are only partially understood, in general, for this type of
iteration \cite{HjalGD:96}.



\psfrag{norm(uf-udes)}{}
\psfrag{om1h}{$
\bea{l} {}\\ \omh_1/\om_1 \eea
$}
\psfrag{nsa=1}{\blue{$\nsa=1$ @ $\tf$}}
\psfrag{nsa=2}{\green{$\nsa=2$ @ $[\tf,\ \tf/2]$}}

\subsection{Direct adaptive control}
\label{sec:direct adapt}

In a direct adaptive control system no model is posed for the system;
ideally only a performance measure is available and the control
parameters are adjusted to improve the performance. The adjustment
``directions'', however, clearly must depend on the shape of the
``landscape'', otherwise, it would not be possible to know how to make
the adjustment. In effect then, a model of the landscape is either
available or is computed intrinsically. Consider the following
bipartite system whose Hamiltonian depends on the two controls
$(\eps_z,\ \eps_x)$.
\[
\bea{rcl}
H(\eps) &=& H_Q(\eps)\otimes I_E+I_Q\otimes H_E + H_{QE}
\\
H_Q(\eps) 
&=&
(\eps_z-1)\om_{Qz}Z/2 + \eps_x\om_{Qx}X/2
\\
H_E &=& \om_{Ez}Z/2 + \om_{Ex}X/2
\\
H_{QE} &=& \om_{QE}
\left(
X^{\otimes 2}+Y^{\otimes 2}+Z^{\otimes 2}
\right)
\eea
\]
The Q-part of the system is assumed to be accessible to the user and
the E-part, the ``environment'', is not.  The goal is to select the
controls to make the Q-system behave as a {\em bit-flip} device, \ie,
the Pauli $X$ matrix. Suppose the Q-system is prepared in the initial
state $\rho_Q^\init=\ket{1}\bra{1}$ and a measurement is made at
$\tf=\pi/\om_{Qx}$ of the state $\ket{0}$. hence, ideally, the outcome
probability, $p(\eps)$, should be unity. Due to the uncontrolled
$E$-system coupling, however, the goal is to select the controls
$\eps$ to make $p(\eps)$ as large as possible. Under these conditions
the outcome probability $p(\eps)$ arises from,
\[
\bea{rcl}
p(\eps) &=& \trace\ M U(\tf)\rho_0 U(\tf)^*
\\
i\dot U(t) &=& H(\eps(t)) U(t),\ U(0) = I,\ 0\leq t\leq\tf
\\
\rho_0 &=& \rho_Q^\init \otimes \rho_E^\init
\eea
\]
Of course the system Hamiltonian is not known, only at best the
outcome probability would be known after enough repetitions of the
experiment.  

\begin{figure}[h]
\centering
\epsfysize=3.5in
\epsfxsize=6.5in
\epsfboxit{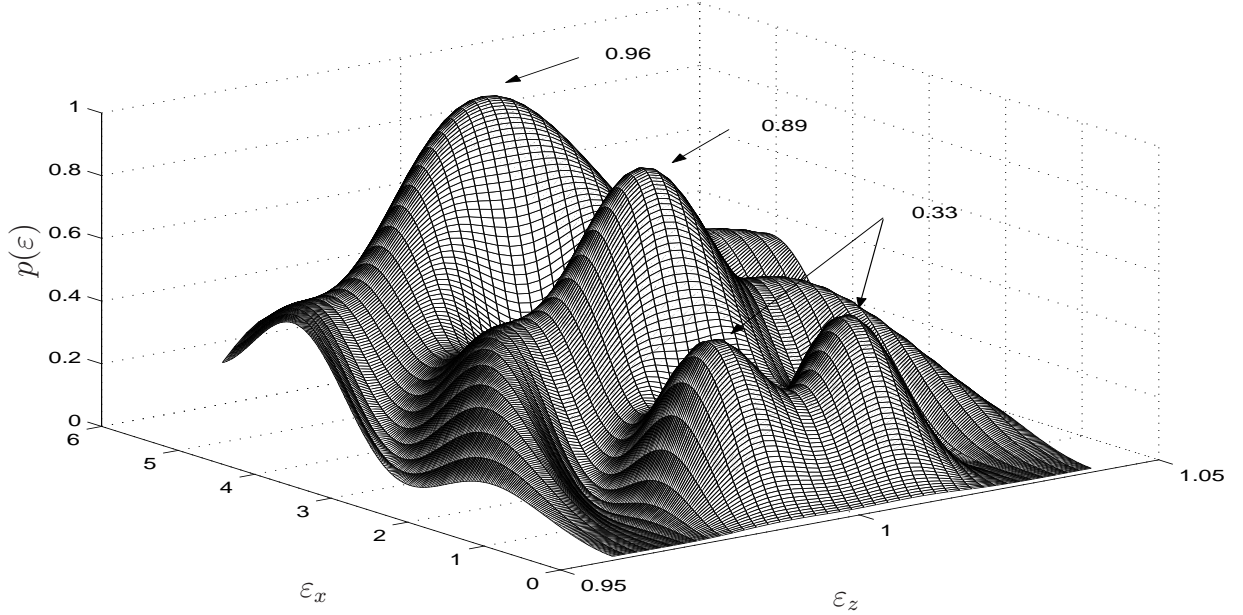}
\caption{Two parameter landscape; the maximum probability
of 0.96 is achieved
with $\eps_z=0$ and $\eps_x=5.2$.}
\label{fig:angel}
\end{figure}

Figure \ref{fig:angel} shows the landscape of the system with
parameters,
$
\om_{Qz}=1,\ \om_{Qx}=0.01,\
\om_{Ez}=1,\ \om_{Qx}=0,\
\om_{QE}=0.005,\
\rho_E^\init=I_2/2
$
and for {\em constant} values of the controls over the ranges $0.96
\leq \eps_z \leq 1.04,\ 0.1 \leq \eps_x \leq 5.2$.


\psfrag{pz}{$p(\eps)$}
\psfrag{ex}{$\eps_z$}
\psfrag{ey}{$\eps_x$}

The landscape clearly has several local maximum values. Thus without
some knowledge of the landscape or an exhaustive search, it might be
difficult to find the global maximum. In this two-parameter case, of
course, an exhaustive search is not too exhausting. Nevertheless, it
is clear that any direct adaptive algorithm will face some
difficulties.  It is also clear that prior knowledge can alleviate
many of the difficulties, \eg, knowledge of system parameter ranges or
nominal response which is close to a good outcome, \etc.  

A more in-depth analysis of the landscape for control of quantum
systems can be found in \cite{RabitzHR:04}. There it is shown that for
unconstrained time-varying controls, if the system is controllable,
then {\em all} the local maximum are global. That is, the outcome
probability at every local maximum is unity and all other extrema give
the minimum probability of zero. The complexity shown here results
principally from the fact that the choice of controls is constrained
to be constant. This points out the importance of constraints either
imposed by the physics or by the designer in inadvertently making the
"wrong" choice of the control structure.  In particular, suppose that
the landscape is actually very simple with even possibly one extremum
when viewed with no constraints on the controls.  However, when
constraints are imposed, or a new set of controls is defined, the
landscape may then exhibit structure that was not evident in the
freely floating original set.  An extreme opposite case could be for a
choice of variables where the new variables actually do hardly
anything at all with regard to control action. In this case one would
conclude that the landscape is totally flat!

Many of the current "working" adaptive feedback control experiments
are operating somewhere between these two extremes, ranging from
having highly constrained knobs to the totally wrong knobs.  This
comment comes from the observation that physical effects spanning a
dynamic range in quantum wavelength of about $10^7$ are being
controlled by a single type of laser, \ie, the Ti:Sapphire laser,
working over a range of about 1-10 on that scale.  That is, a domain
in laser wavelength of 10 (in some units) is split up into about 100
small pieces as controls, and those very narrow controls manage
everything over a huge dynamic range. The fact that the experiments
work at all is rather amazing. A guess is that an examination of the
landscapes will show considerable detail, much of which is likely
false structures arising from having highly constrained controls.
From a positive perspective, as more bandwidth becomes available the
control landscape will become less complicated and more regular in the
sense that more of the local optimum values will provide performance
close to the global optimum.


\begin{appendix}
\section{Appendix}
\subsection{Worst-case gate fidelity}
\label{sec:fidelity}

From the definition of worst-case fidelity \refeq{wc fidelity},
\[
\bea{rcl}
\left|
\left(\Udes \psi\right)^*
\left( U_{\rm act}\psi \right)
\right|^2
&=&
\left|
\psi^* (\Udes^*\Uact)\psi
\right|^2
\\
&=&
\left|
(V^*\psi)^*\Om(V^*\psi)
\right|^2
\\
&=&
\left|
\sum_{k=1}^n\ \om_k\ |x_k|^2
\right|^2
\\
&=&
\left|
\sum_{k=1}^n\ \om_k\ z_k
\right|^2
\eea
\]
where the last three lines follow directly from (i) the eigenvalue
decomposition of the unitary: $\Udes^*\Uact=V\Om V^*,\ V^*V=I_n,\
\Om=\diag(\om_1\ \cdots\ \om_n),\ |\om_k|=1$, (ii) defining
$x=V^*\psi\in\Cbf^n$, and (iii) defining $z_k=|x_k|^2\in\Rbf$. Using
the definitions of the vectors $(a,b)$ in \refeq{fidelity qp} gives,
\[
\left|
\left(\Udes \psi\right)^*
\left( U_{\rm act}\psi \right)
\right|^2
=
\left|
\sum_{k=1}^n\ (a_k+ib_k)\ z_k
\right|^2
=
z^T(aa^T+bb^T)z
\]
The QP follows from the relations:
\[
\norm{\psi}=1\
\Leftrightarrow\
\norm{x=V^*\psi}=1\
\Leftrightarrow\
\sum_k\ (z_k=|x_k|^2) = 1,\ z_k\geq 0
\]
%

\subsection{\crao Inequality}
\label{sec:crao}

The following is the classical form of the \crao Inequality.

\bquote
{\bf Cram\'{e}r-Rao Inequality}\cite{Cramer:46}
\\
Let $\th_0\in\Rbf^p$ be the true parameter to be estimated from a data
set $D$. Let $L(D,\th_0)$ be the true negative log-likelihood function
of the system generating the data. Let $\thh\in\Rbf^p$ be an unbiased
estimate of $\th_0$, \ie, $\avg\ \thh=\th_0$. Then, the covariance of
the estimate,
\beq[eq:covth]
\cov{\thh} 
= 
\avg{
\left(
\thh-\th_0
\right)
\left(
\thh-\th_0
\right)^T
}
\eeq
satisfies the matrix inequality,
\beq[eq:cramer1]
\left[
\bea{cc}
\cov\ \thh & I
\\
I & F(\th_0)
\eea
\right]
\geq 0
\eeq
where $F(\th_0)$ is the {\em Fisher information matrix},
\beq[eq:fisher]
F(\th_0) 
=
\avg\
\nabla_{\th\th}
L(D,\th)
\bigg|_{\th=\th_0}
\eeq
If $F(\th_0)>0$, then \refeq{cramer1} is equivalent to,
\beq[eq:cramer2]
\cov{\thh}
\geq
F(\th_0)^{-1}
\eeq
\equote
This famous theorem states the for any unbiased estimator, the
covariance of the estimate satisfies the inequality \refeq{cramer1},
or equivalently \refeq{cramer2}, provided the Fisher matrix is
invertible. Usually only \refeq{cramer2} is given as the theorem: the
minimum covariance of the estimate is given by the inverse of the
Fisher information matrix. The power of the result \refeq{cramer1} is
that it is independent of {\em how} the estimate is obtained. The
lower bound only depends on the model structure, the experiment
design, and the information in the data.

\subsection{Derivation of \refeq{var rho}}
\label{sec:var rho}

Define the vectors $r,\ a_{\alf\gam}\in\Cbf^{n^2}$ as,
\[
r=\vec\ \rho,
\;\;\;
a_{\alf\gam}=\vec\ O_{\alf\gam}
\]
Then 
\[
p_{\alf\gam}=\trace\ O_{\alf\gam}\rho = a_{\alf\gam}^* r
\]
and
\[
\bea{rcl}
\trace\ \rho=1
&\Leftrightarrow&
b^T r=1,
\;\;
b=\vec\ I_n
\eea
\]
The next step eliminates the equality constraint $b^Tr=1$ reducing the
$n^2$ unknowns in $r$ to $n^2-1$.  Since $b^Tb=n$, the SVD of $b$ is,
\[
b=W\left[\bea{c} \sqrt{n} \\ 0_{n^2-n} \eea \right]
\]
with unitary $W\in\Rbf^{n^2\times n^2}$, Partition $W=[c\ \Ceq]$ with
$\Ceq\in\Rbf^{n^2\times n^2-1}$. Then all $r$ satisfying $b^Tr=1$ are
given by
\[
r=c/\sqrt{n} + \Ceq z
\]
for all $z\in\Cbf^{n^2-1}$. The likelihood function with the equality
constraint ($\trace\ \rho=1$ or $b^Tr=1$) eliminated is then a function
only of $z$,
\[
L(D,z)=-\sum_{\alf,\gam} n_{\alf\gam} \log (c/\sqrt{n} + \Ceq z)
\]
To obtain the \crao bound, we first compute,
\[
\nabla_{zz}L(D,z)
=
\sum_{\alf,\gam} 
\frac{n_{\alf\gam}}{p_{\alf\gam}(z)^2} 
(\Ceq^T a_{\alf\gam})(\Ceq^T a_{\alf\gam})^*
\]
Using \refeq{av nalfgam}, $\avg\ n_{\alf\gam}=\ell_\gam
p_{\alf\gam}(\rhotrue),\ \rhotrue=c/\sqrt{n}+\Ceq z^{\rm true}$, and hence
the Fisher information matrix is, with respect to $z$,
\[
\bea{rcl}
F &=& \avlike(D,z=z^{\rm true})
\\
&=&
\ds
\sum_{\alf,\gam} 
\frac{\ell_\gam}{p_{\alf\gam}(\rhotrue)} 
(\Ceq^T a_{\alf\gam})(\Ceq^T a_{\alf\gam})^*
\\
&=&
G(\ell,\rhotrue)
\eea
\]
where the last line comes from the definition of $G(\ell,\rhotrue)$ in
\refeq{var rho 1}. Let $\rh=c/\sqrt{n}+\Ceq\zh$ be an unbiased
estimate of $r^{\rm true}$. Thus,
\[
\cov\  \rh = \Ceq \cov\  \zh\ \Ceq^T \geq \Ceq F^{-1} \Ceq^T 
\]
Using $\var\  \rhoh = \var\  \rh =\trace\ \cov\  \rh$ and the fact that $W$ is
unitary, and hence, $\Ceq^T\Ceq=I_{n^2-1}$, we get,
\[
\var\  \rhoh \geq \trace\ \Ceq F^{-1} \Ceq^T = \trace\ F^{-1}
\]
which is the final result \refeq{var rho}-\refeq{var rho 1}.

\subsection{Derivation of \refeq{expdes osr}}
\label{sec:expdes osr der}

Define $x,\ r_{\alf\gam}\ \in\Cbf^{n^4}$ as,
\[
x = \vec\ X,
\;\;
r_{\alf\gam} = \vec\ R_{\alf\gam}
\]
The likelihood function in \refeq{mle x} can then be written as,
\[
L(D,x)=-\sum_{\alf,\gam}\ n_{\alf\gam} \log r_{\alf\gam}^* x
\]
and the equality constraint in \refeq{mle x} becomes,
\[
Ax=\vec\ I_n,
\;\;
A=[a_1\ \cdots\ a_{n^4}]\in\Cbf^{n^2 \times n^4}
\]
with $a_k=\vec(B_i^*B_j)\in\Cbf^{n^2}$ for $k=i+(j-1)n^2,\
i,j=1,\ldots,n^2$.  Perform the singular value decomposition
$A=U[S\ 0]W^*,\ W=[C\ \Ceq]$ with $C\in\Cbf^{n^2\times n^2},\,
\Ceq\in\Cbf^{n^2\times n^4-n^2}$ as given in \refeq{expdes osr 2}.
From the definition of the basis functions \refeq{basis} it follows
that $S=\sqrt{n}\ I_{n^2}$.  Observe also that the columns of $\Ceq$
are a basis for the nullspace of $A$. Hence all $x$ satisfying the
equality constraint are given by,
\[
x=\bar{x}+\Ceq z,\; \forall z
\;
\mbox{with}
\;
\bar{x}=(1/\sqrt{n}) C U^*\vec\ I_n
\]
The likelihood function with the equality constraint ($Ax=\vec\ I_n$)
eliminated is then a function only of $z$,
\[
L=-\sum_{\alf,\gam} n_{\alf\gam} \log (\bar{x} + \Ceq z)
\]
This is exactly the same form of the likelihood function in
\refsec{var rho} after the single equality constraint there is
eliminated. Hence, to obtain the \crao bound \refeq{expdes osr}
repeat, {\em mutatis mutandis}, the procedure in \refsec{var rho}.

\end{appendix}

\bibliographystyle{plain}
\bibliography{D:/robert/tex/rlk}

\end{document}

%% file: articlelayout.tex
\newlength{\medpaperheight}     
\newlength{\medpaperwidth}      
\newlength{\medtextheight}      
\newlength{\medtextwidth}       
\newlength{\medtopmargin}       
\newlength{\medoddsidemargin}   

%% file: rlktexdefs.tex
\newcommand{\real}{{\rm Re}}
\newcommand{\imag}{{\rm Im}}
\newcommand{\diag}{{\rm diag}}

\newcommand{\eps}{\varepsilon}

\newcommand{\gam}{\gamma}
\newcommand{\alf}{\alpha}
\newcommand{\om}{\omega}
\newcommand{\lam}{\lambda}
\newcommand{\bet}{\beta}
\renewcommand{\th}{\theta}
\newcommand{\del}{\delta}
\newcommand{\sig}{\sigma}

\newcommand{\Del}{\Delta}

\newcommand{\Om}{\Omega}

\newcommand{\norm}[1]{ \left\| #1 \right\| }

\newcommand{\lamh}{\hat{\lam}}

\newcommand{\rhoh}{\hat{\rho}}
\newcommand{\rh}{\hat{r}}

\newcommand{\thh}{{\hat{\theta}}}

\newcommand{\epsb}{\bar{\eps}}

\newcommand{\fracds}[2]{\frac{\ds #1}{\ds #2}}

\newcommand{\Astrom}{\r{A}str\"{o}m~}

\newcommand{\eg}{\emph{e.g.}}

\newcommand{\ie}{\emph{i.e.}}
\newcommand{\etc}{\emph{etc.}}

\newcommand{\bquem}{\begin{quote}\begin{em}}
\newcommand{\equem}{\end{em}\end{quote}}

\newcommand{\blist}{\begin{description}}
\newcommand{\elist}{\end{description}}

\newcommand{\bquote}{\begin{quote}}
\newcommand{\equote}{\end{quote}}

\newcommand{\ben}{\begin{enumerate}}
\newcommand{\een}{\end{enumerate}}

\newcommand{\bit}{\begin{itemize}}
\newcommand{\eit}{\end{itemize}}

\newcommand{\bea}{\begin{array}}
\newcommand{\eea}{\end{array}}

\newcommand{\bds}{\begin{displaystyle}}
\newcommand{\eds}{\end{displaystyle}}

\newcommand{\Rbf}{{\mathbf R}}
\newcommand{\Cbf}{{\mathbf C}}

\newcommand{\ds}{\displaystyle}

\newcommand{\fs}{\footnotesize}

\newcommand{\mbf}[1]{\mbox{\boldmath $#1$}}

\newcommand{\refeq}[1]{(\ref{eq:#1})}
\newcommand{\reffig}[1]{\ref{fig:#1}}
\newcommand{\refsec}[1]{\ref{sec:#1}}

\newcommand{\set}[2]{ \left\{ \,#1\, \left| \,#2\, \right.\right\} }

\newcommand{\seq}[1]{ \left\{ #1 \right\} }

\newcommand{\sett}[2]{ \Big\{ \,#1\, \Big| \,#2\, \Big\} }

\newcommand{\opt}{{\rm opt}}
\newcommand{\expt}{{\rm expt}}

\newcommand{\qt}[1]{``#1''}

\newcommand{\mathbox}[1]{
\fbox{$\ds #1 $}
}

\makeatletter
\def\beq{\@ifnextchar 
[{\@tempswatrue\@beq}{\@tempswafalse\@beq[]}}
\def\@beq[#1]{\begin{equation}\edef\@tmparg{#1}\ifx\@tmparg\@e
mpty \else
	\label{#1}\fi}
\newcommand{\eeq}{\end{equation}}
\makeatother 

\newcommand{\beqaa}{\begin{eqnarray*}}
\newcommand{\eeqaa}{\end{eqnarray*}}

\newcommand{\beqa}{\begin{eqnarray}}
\newcommand{\eeqa}{\end{eqnarray}}

\newcommand{\bc}{\begin{center}}
\newcommand{\ec}{\end{center}}

